\documentclass[12pt]{article}
\usepackage{amsfonts,amsmath,amssymb,epsfig}

\makeatletter\@addtoreset{equation}{section}\makeatother

\def\bR {\mathbb{R}}
\def\bZ {\mathbb{Z}}

\input epsf

\newcommand{\vev}[1]{{\left< {#1} \right>}}

\newcommand{\Tr}{{\rm Tr\,}}

\newcommand{\cN}{{\cal N}}
\newcommand{\cP}{{\cal P}}

\newcommand{\cS}{{\cal S}}

\newcommand{\preprint}[1]{\begin{table}[t]  %%
             \begin{flushright}               %%
             {#1}                             %%
             \end{flushright}                 %%
             \end{table}}                     %%
\renewcommand{\title}[1]{\vbox{\center\LARGE{#1}}\vspace{5mm}}
\renewcommand{\author}[1]{\vbox{\center#1}\vspace{5mm}}
\newcommand{\address}[1]{\vbox{\center\em#1}}
\newcommand{\email}[1]{\vbox{\center\tt#1}\vspace{5mm}}

\begin{document}

\begin{titlepage}
\preprint{hep-th/0506058 \\ 
ITFA-2005-19
}

\title{On the integrability of Wilson loops in $AdS_5\times S^5$:\\
Some periodic ansatze}

\author{Nadav Drukker$^1$ and Bartomeu Fiol$^2$}

\address{$^1$The Niels Bohr Institute, Copenhagen University\\
Blegdamsvej 17, DK-2100 Copenhagen, Denmark\\
\smallskip
$^2$Institute for Theoretical Physics, University of Amsterdam\\
1018 XE Amsterdam, the Netherlands
}

\email{drukker@nbi.dk, bfiol@science.uva.nl}

\abstract{
Wilson loops are calculated within the $AdS$/CFT correspondence by 
finding a classical solution to the string equations of motion in 
$AdS_5\times S^5$ and evaluating its action. An important fact is 
that this $\sigma$-model used to evaluate the Wilson loops is 
integrable, a feature that has gained relevance through the study of 
spinning strings carrying large quantum numbers and spin-chains. 
We apply the same techniques used to solve the equations for spinning 
strings to find the minimal surfaces describing a wide class of Wilson 
loops. We focus on different cases with periodic boundary conditions 
on the $AdS_5$ and $S^5$ factors and find a rich array of solutions. 
We examine the different phases  that appear in the problem and 
comment on the applicability of integrability to the general problem.}

\end{titlepage}

\section{Introduction}

Wilson loops are some of the most interesting observables in non-Abelian 
gauge theories. In addition to their paramount role as the order parameter 
for confinement, they constitute a large class of non-local gauge invariant 
observables. In the $AdS$/CFT \cite{ads/cft} 
correspondence they are realized in a very suggestive way: 
the expectation value of the Wilson loop is given by the string partition 
function with appropriate boundary conditions. The superstring in 
$AdS_5\times S^5$ bound by the loop represents the Wilson loop 
in the same way the QCD string should give the area-law of a confining 
gauge theory.

In practical applications one uses $AdS_5\times S^5$ to evaluate the 
Wilson loop at large~$N$ and 't Hooft coupling $\lambda=g_{YM}^2N$. 
On the string theory side this translates to weak string coupling 
$g_s=\lambda/(4\pi N)$ and large string tension (or large $AdS_5$ and 
$S^5$ radii $L$), since $\lambda=L^4/\alpha'^2$. In this limit the 
string calculation reduces to finding the minimal surface with 
appropriate boundary conditions and evaluating its action.

Despite their importance, the study of Wilson loops in the $AdS$/CFT
correspondence hasn't benefited from the same impetus as that of
local gauge theory operators. In particular, the exciting discovery of 
integrable structures on both sides of the duality has been so far applied 
only to local operators. One of the main purposes of the present 
work is to extend the tools of integrability also to operators that are 
non-local on the gauge theory. 

In order to place this work in perspective, it is useful to remind ourselves
of the status of classical integrability in $AdS$/CFT, mostly on the string 
theory side of the duality (for reviews on the impressive body of work on 
integrability on the gauge theory side, see e.g. \cite{Beisert:2004ry}). 
Integrability of two-dimensional non-linear sigma models has been 
known for many years \cite{Pohlmeyer:1975nb}. The world-sheet 
description of 
string theory on $AdS_5 \times S^5$ is given by two non-linear sigma models, 
related by the Virasoro constraints \cite{Metsaev:1998it}. In 
\cite{Mandal:2002fs} it was observed that the 
presence of Virasoro constraints doesn't affect the construction of currents, 
so the full bosonic sector of the world-sheet description is integrable. It 
was later shown in \cite{Bena:2003wd} that classical integrability extends 
to the fermionic sector.

These non-linear sigma models can be simplified by considering particular 
ansatze. A prime example are the spinning string solutions 
\cite{Gubser:2002tv, Frolov:2002av}. In the world-sheet this ansatz reduces 
the full non-lineal $\sigma$-model to a much simpler 1$d$ integrable system 
\cite{Arutyunov:2003uj, Arutyunov:2003za}.

The simple but far-reaching observation, which is the starting point of this 
work, is that locally supersymmetric 
Wilson loops are calculated within the $AdS$/CFT 
correspondence using the same integrable classical $\sigma$-model. 
While for the local operators the classical description in terms of string 
solutions is valid only at large quantum numbers, the Wilson loops we consider 
may always be studied by semiclassical methods. Thus the integrability of 
the $\sigma$-model should be utilized to evaluate those non-local 
observables.

In this paper we perform some concrete calculations of Wilson loop 
observables by considering periodic ansatze similar to those of the rotating 
strings \cite{Arutyunov:2003uj, Arutyunov:2003za}. The resulting 
surfaces will evaluate Wilson loops whose contours are symmetric, like 
straight lines and circles, and with periodic couplings to the scalar fields.
Those will include as particular cases many of the Wilson loops evaluated 
previously within the $AdS$/CFT framework.

Note that we will study the Wilson loops only on the string side of the 
duality. For the local operators, an integrable structure was also 
found in the gauge theory at weak coupling in terms of the Bethe 
ansatz solution of spin-chains \cite{Minahan:2002ve}. 
In certain cases the correspondence with spinning 
strings allows to calculate conformal dimensions reliably on both sides 
and compare the results, which exhibit remarkable agreement. 
But the statement about integrability of the $\sigma$-model 
does not require this agreement between strong and weak coupling. In fact, 
in the examples we study, we generally did not find a simple agreement 
between the gauge theory calculation and the one in 
$AdS_5\times S^5$, so we are focusing exclusively on the latter.

One of the most salient aspects of the present work is that 
the geometric data that define the Wilson loop operator (e.g. distance 
between parallel lines, ratio of radii), is not all encoded in a 
straight-forward 
manner in terms of the parameters and the integrals of motion 
of the 1$d$ integrable model. In the general case we will see 
that they are related by equations that can't be easily inverted, and require 
numerical analysis. More than that, there is not even a one to one 
correspondence and generically there are many classical solutions to the 
$\sigma$-model for the same boundary conditions.

As stated above, in this paper we will focus on the string theory 
side of the duality, and already there we will find some very confusing 
phenomena. We will comment on some of the difficulties in the 
comparison to the weakly coupled gauge theory in the final section. 

It is worth mentioning the one case where there is good agreement 
between the gauge theory and string results, which is the circular 
Wilson loop (with simplest coupling to the scalars)
\cite{Berenstein:1999ij,Drukker:1999zq}. In that case 
summing up the perturbative series (assuming no interactions) leads 
exactly to the string result including all non-planar corrections
\cite{erickson,Drukker:2000rr,Drukker:2005kx}. 
The agreement seems to hold for both the expectation value of the 
Wilson loop, as well as the correlation function of the Wilson loop 
with some local operators 
\cite{Semenoff:2001xp,Zarembo:2002ph}.

The paper is organized as follows. In the rest of the introduction we 
present the types of Wilson loops we will be studying. Then we will 
set up the $\sigma$-model calculation. In Section~\ref{S5-section} 
we look at the $S^5$ part of the equations and solve them for periodic 
motion. We write explicitly the solutions for motions in $S^1$, $S^2$ 
and $S^3$ subspaces which we utilize later. In the following section 
we do the same for the $AdS_5$ part of the action, solving it for 
general periodic motions and then concentrating on solutions within 
$AdS_2$ and $AdS_3$ subspaces. Those include a line, circle, parallel 
lines and concentric circles.

In Section~\ref{classification-section} we put this all together into 
full solutions in $AdS_5\times S^5$. We start with solutions in 
$AdS_2$, which are just the straight line and the circle. This is the one 
case where there is an explicit calculation relating the string results to 
the gauge theory, and we review it there. We then discuss solutions 
within $AdS_3\times S^1$, which include two lines or concentric 
circles that may be separated on the $S^5$. The case of the lines 
was already studied in the original papers 
\cite{rey-wl,maldacena-wl} and the circles (without the $S^1$ 
dependence) were studied before in 
\cite{Zarembo:1999bu,Olesen:2000ji}.

The next solutions we consider live in an $AdS_2\times S^2$ subspace 
of $AdS_5\times S_5$. They correspond to a line or circle with 
periodic motion inside an $S^2$, generalizing solutions presented 
in \cite{Zarembo:2002an,Tseytlin:2002tr}. Finally we study the 
periodic motions inside an $AdS_3\times S^3$ subspace (the 
simplest of those was discussed in \cite{Tseytlin:2002tr}). Those 
are parallel lines and concentric circles that also rotate on an $S^2$, 
and they exhibit very rich phenomena. We find at least four types 
of classical solutions and describe the phase transitions between them. 

We end with a discussion on the results and some general speculations 
on the integrability of Wilson loop operators.

\subsection{The observables}

The (locally) supersymmetric Wilson loops in $\cN=4$ supersymmetric 
Yang-Mills include a coupling to both the gauge field $A_\mu$ and 
the scalars $\Phi_i$. The general observable is
\begin{equation}
W=\frac{1}{N}\Tr \cP
\exp\left[i\int \left(A_\mu\dot x^\mu(t)
+i|\dot x(t)|\Theta^i(t) \Phi_i\right)dt\right]\,.
\end{equation}
Here $x^\mu$ describes the curve in space and $\Theta^i$ are 
unit vectors in $\bR^6$. More general operators will include also 
couplings to the fermionic fields, but we will study the bosonic 
operators only.

The fact that the magnitude of the coupling to the scalar is equal to 
the coupling to the gauge field is crucial to guarantee the existence 
of a classical solution to the equations of motion. It is actually not 
know how to evaluate Wilson loops not obeying this constraint. 
Another point to note is the extra factor of $i$ in front of the scalar 
term. This shows up upon Wick rotation to the Euclidean theory 
(which we study here), and is also natural for spatial loops in the 
Lorentzian case.

We will be interested in loops following circular or straight 
paths in space. First we shall look at the infinite line
\begin{equation}
x^1=t\,,
\label{line-loop}
\end{equation}
and then at the circle
\begin{equation}
x^1=R\cos kt\,,\qquad
x^2=R\sin kt\,.
\label{circle-loop}
\end{equation}
In both cases we will label the length of the line by $T$, so $0<t<T$. 
In the first case $T$ will be a cutoff on the diverging length, and in the 
second case $T=2\pi$. Then the integer $k$ describes the number of 
times the loop wraps the circle. We will also consider 
the correlator of two of those operators.

While we will not study those cases in detail, we will also solve the 
equations of motion (implicitly) for loop with rotation in two planes
\begin{equation}
\begin{gathered}
x^1=R_1\cos k_1t\,,\qquad
x^2=R_1\sin k_1t\,,\\
x^3=R_2\sin k_2t\,,\qquad
x^4=R_2\sin k_2t\,,
\end{gathered}
\end{equation}
with $k_1\neq k_2$ two integers. There is also the helical Wilson loop, 
where one of the circles is replaced with a straight line ($R_2$ taken 
to be infinite)
\begin{equation}
x^1=R\cos k_1t\,,\qquad
x^2=R\sin k_1t\,,\qquad
x^3=t\,,
\end{equation}

Our general ansatz will address the possibility of turning on as many as all 
six of the scalar fields in the following way
\begin{equation}
\begin{aligned}
&\Theta^1(t)+i\Theta^2(t)
=\sin \theta\,e^{i(m_1t+\varphi_{1})}\,,\\
&\Theta^3(t)+i\Theta^4(t)
=\cos \theta\sin\psi\,e^{i(m_2t+\varphi_{2})}\,,\\
&\Theta^5(t)+i\Theta^6(t)
=\cos \theta\cos\psi\,e^{i(m_3t+\varphi_{3})}\,.
\label{S5-loop}
\end{aligned}
\end{equation}
In the example we study in detail we will turn on at most three 
scalars, i.e. the above ansatz with $m_2=m_3=\cos\psi=\varphi_2=0$

If $\theta=\pi/2$ and $m_1=0$ there is only a 
coupling to $\Theta^1$, which is the case considered most often. 
But some more complicated couplings to the sphere were considered 
in the past, for example, $\theta=\pi/2$ and $m_1=k$ is the 
supersymmetric loop studied in \cite{Zarembo:2002an}.

The map from the gauge theory 
to $AdS_5\times S^5$ is the following: the Wilson loop will be 
described by a minimal surface that will 
follow the contour $x^\mu$ on the boundary, and the derivative 
normal to the boundary of the surface of the radial coordinate $y$ and 
angular coordinates, combined together into a six-vector, is 
proportional to $\Theta^i$, so
\begin{equation}
\partial_\sigma (y^i)\propto\Theta^i\,.
\end{equation}
In all the cases we study the surfaces are smooth, which implies that 
$\Theta^i(t)$ will be equal to the boundary values of the $S^5$ 
coordinates.

\subsection{The $\sigma$-model}

The bosonic part of the action of a string in $AdS_5\times S^5$ is 
a standard $\sigma$-model
\begin{equation}
\frac{1}{4\pi\alpha'}\int d\tau\,d\sigma\,
\sqrt{g}\,g^{\alpha\beta}
\partial_\alpha X^M\partial_\beta X^N G_{MN}\,.
\end{equation}
Here $G_{MN}$ is the target space metric for $AdS_5\times S^5$ each 
with curvature radius $L$. By the $AdS$/CFT correspondence it is 
related to the 't Hooft coupling $\lambda=g_{YM}^2N$ of the dual 
gauge theory and the string scale by $L^4=\lambda\alpha'^2$.

The ansatz we consider factorizes into an $AdS_5$ part and an $S^5$ part, 
yielding 
independent equations of motion for the respective variables. These two 
parts of the ansatz are related only in two ways. One is the range of the 
world-sheet coordinates, which clearly has to agree, and the other are the 
Virasoro constraints. Therefore, it makes sense to consider separately the 
$S^5$ and $AdS_5$ parts of the ansatz.

The Virasoro constraints are the vanishing of the stress-energy 
tensor which in the conformal gauge is given by
\begin{equation}
\begin{gathered}
{T_{\sigma\sigma}=-T_{\tau\tau}
=\frac{1}{8\pi\alpha'}\left[
\partial_\sigma X^M\partial_\sigma X^N 
-\partial_\tau X^M\partial_\tau X^N 
\right]G_{MN}=0\,,}\\
T_{\sigma\tau}=T_{\tau\sigma}
=\frac{1}{4\pi\alpha'}\,
\partial_\sigma X^M\partial_\tau X^NG_{MN}=0\,.
\end{gathered}
\end{equation}

Since our space has a product structure we can decompose the 
stress-energy tensor into independent contributions from $AdS_5$ 
and from $S^5$
\begin{equation}
T_{\alpha\beta}
=T_{\alpha\beta}^{AdS_5}+T_{\alpha\beta}^{S^5}\,.
\end{equation}
The Virasoro constraints are then
\begin{equation}
T_{\alpha\beta}^{AdS_5}+T_{\alpha\beta}^{S^5}=0\,.
\end{equation}
For notational simplicity we label 
$a^2\equiv8\pi\alpha'T_{\sigma\sigma}^{S^5}/L^2$, 
and in $AdS_5$ this parameter serves a role similar to a mass term 
coming from the Kaluza-Klein reduction on the sphere. 

Since the stress-energy tensors
of both $\sigma$-models are separately conserved, we have
\begin{equation}
\partial_\sigma T^{S^5}_{\sigma \sigma}
+\partial_\tau T^{S^5}_{\tau \sigma}=0
\end{equation}
and a similar equation for $AdS_5$. In the ansatze we will use below, 
$T^{S^5}_{\sigma \tau}$ is always constant (actually zero for the examples
we consider in more detail), so it follows that $a^2$ is 
constant. Note that $a^2$ may be either positive or negative.

\section{The $S^5$ ansatz}
\label{S5-section}

We start by studying the $\sigma$-model on $S^5$. The standard metric 
on the sphere is
\begin{equation}
ds^2_{S^5}=L^2\left(d\theta ^2+\sin^2\theta d\varphi_1^2
+\cos^2\theta\left(d\psi^2+\sin^2\psi d\varphi_2^2
+\cos^2\psi d\varphi_3^2\right)\right)\,.
\end{equation}
Following \cite{Arutyunov:2003uj,Tseytlin:2003ii}, we use embedding 
coordinates in flat $\bR^6$ by defining three radial coordinates
\begin{equation}
\rho_1=\sin\theta\,,\qquad
\rho_2=\cos\theta\sin\psi\,,\qquad
\rho_3=\cos\theta\cos\psi\,,
\end{equation}
which clearly satisfy $\sum \rho_i^2=1$, which we impose through 
the inclusion of a Lagrange multiplier $\Lambda$.
The $S^5$ part of the action is
\begin{equation}
\cS_{S^5}=\frac{L^2}{4\pi\alpha'}
\int d\sigma\,d\tau \Bigg[\sum_{i=1}^3\left(\rho_i'^2+\dot \rho_i^2
+\rho_i^2(\varphi_i'^2+\dot\varphi_i^2)\right)
+\Lambda\left(\sum_{i=1}^3 \rho_i^2-1\right)\Bigg]\,.
\end{equation}

We are interested in the ansatz\footnote{
Compared to the spinning string ansatz the coordinates $\sigma$ and 
$\tau$ are reversed, since we consider the direction along the curve at 
the boundary to be the (Euclidean) time direction.}
\begin{equation}
\rho_i=\rho_i(\sigma)\,,\qquad
\varphi_i=m_i\tau+\beta_i(\sigma)\,,
\label{S5-ansatz}
\end{equation}
where $m_i$ are arbitrary constants (which have to be integers for 
a compact world-sheet). One 
can check that this ansatz solves the equations of motion and 
it leads to the action
\begin{equation}
\cS_{S^5}=\frac{L^2}{4\pi\alpha'}
\int d\sigma\,d\tau  \Bigg[\sum_{i=1}^3\left(\rho_i'^2
+\rho_i^2(\beta_i'^2+m_i^2)\right)
+\Lambda\left(\sum_{i=1}^3 \rho_i^2-1\right)\Bigg]\,.
\label{S5-action}
\end{equation}
The $\tau$ integration just gives an overall factor of the length of 
the Wilson line, $T$, thus we are left with a one-dimensional problem, the 
Neumann-Rosochatius system, a particular case of the $n=6$ Neumann system.
Clearly $\beta_i$ are cyclic, 
with conserved momenta $\pi_i$, so
\begin{equation}
\beta_i'=\frac{\pi_i}{\rho_i^2}\,.
\label{pi's}
\end{equation}
The other equations of motion are
\begin{equation}
\rho_i''=\frac{\pi_i^2}{\rho_i^3}+\rho_i(m_i^2+\Lambda)\,,
\end{equation}

The Neumann-Rosochatius system is integrable, and it's easy to check 
that the following three quantities 
\begin{equation}
I_i=\rho_i^2-\sum_{j\neq i}\frac{1}{m_i^2-m_j^2}
\left((\rho_i\rho_j'-\rho_j\rho_i')^2
+\frac{\pi_i^2}{\rho_i^2}\rho_j^2
+\frac{\pi_j^2}{\rho_j^2}\rho_i^2\right)\,,
\end{equation}
with $i=1,\,2,\,3$ are constants of motion. They are not independent, 
but satisfy $I_1+I_2+I_3=1$. Another identity relates them to the 
$S^5$ contribution to the stress-energy tensor
\begin{equation}
\frac{8\pi\alpha'}{L^2}\,T_{\sigma\sigma}^{S^5}=a^2
=\sum_{i=1}^3\left(\pi_i^2-m_i^2I_i\right)
=\sum_{i=1}^3\left(\rho_i'^2
+\frac{\pi_i^2}{\rho_i^2}-\rho_i^2m_i^2\right)\,.
\label{S5-first-integral}
\end{equation}

The off-diagonal contribution to the stress-energy tensor,
$T_{\sigma\tau}=(L^2/4\pi\alpha')\sum m_i\pi_i$ is also a constant.

When evaluating the classical action of our solutions we may use this 
conserved quantity to replace the potential terms with the kinetic terms
\begin{equation}
\cS_{S^5}
=\frac{\sqrt\lambda}{4\pi}\int d\sigma\,d\tau
\left[2\sum_{i=1}^{3}m_i^2\rho_i^2+a^2\right]
=2\cS_{S^5}^{kinetic}
+\frac{\sqrt\lambda}{4\pi}a^2\delta\sigma T\,.
\label{S5-kinetic}
\end{equation}
$T$ is the range of the coordinate $\tau$, or the length of the Wilson 
loop, and $\delta\sigma$ is the range of the $\sigma$ coordinate. 
So the second term on the right hand side is proportional to the volume of 
the world-sheet. Due to the Virasoro constraint, 
it will be canceled by a similar term from the $AdS_5$ 
part of the $\sigma$-model, the total classical action will be just twice 
the sum of the kinetic terms on both sides. 

Finally, it is worth noticing that since the integrand in (\ref{S5-action}) 
is positive, the $S^5$ contribution to the full action is 
positive, and the same is true for the kinetic term alone.

\subsection{$S^1$ ansatz}
\label{S1-section}

We look now at specific examples, starting with an ansatz that turns 
on only a single angle $\varphi_1$.

In order for the other angles to be constants we have to take 
$\rho_1=1$, or $\theta=\pi/2$ and consequently 
$\rho_2=\rho_3=0$. To turn on this angle in the framework of our 
general ansatz we should take only $\pi_1\neq0$ 
while $\pi_2=\pi_3=0$. We also set all three $m_i=0$. 

There is another possibility, where we turn on $m_1$ instead of $\pi_1$. 
This imposes boundary conditions within this $S^1$ equator of $S^5$, 
but as we will see in the next subsection, the resulting minimal surface 
will generally not stay at $\rho_1=1$. The solution will break the 
symmetry and extend into some other direction, so we treat that case there.

The only equation of motion is $\varphi_1'=\pi_1$, and is solved by 
$\varphi_1=\varphi_{1i}+\pi_1\sigma$. 
The contribution to the stress-energy tensor is proportional 
to $a^2=\pi_1^2$, which will feed into the Virasoro constraint equation. 
The integration constants $I_i$ are not well defined.

This ansatz will be used below when looking at the correlator of two 
Wilson loops separated in the $\varphi_1$ direction as was originally 
studied in \cite{maldacena-wl}. The boundary conditions specifying 
$\varphi_1$ along each loop, $\varphi_{1i}$ and $\varphi_{1f}$, 
will fix $\pi_1$. The difference in the value of this angle 
between the two loops is related to the range of the world-sheet 
variable $\sigma$ by
\begin{equation}
\delta\varphi_1=\pm a\delta\sigma\,.
\label{S1-delta-sigma}
\end{equation}
See section~\ref{AdS3xS1-section} below.

The classical action for this solution has no kinetic term, so when 
combining this ansatz with the $AdS_5$ part the full action will be 
given by twice the kinetic term in the $AdS_5$ action.

\subsection{$S^2$ ansatz}
\label{S2-section}

Let us look now at the case when we turn on two of the angles, 
$\theta$ and $\varphi_1$, by including a single rotation, $m_1=m$. 
Then $\rho_1=\sin\theta$, $\rho_2=\cos\theta$, $\rho_3=0$. 
If we take $\pi_i=0$, the conserved energy (\ref{S5-first-integral}) 
reads
\begin{equation}
\rho_1'^2+\rho_2'^2-m^2\rho_1^2
=\theta'^2-m^2\sin^2\theta
=a^2\,.
\label{S2-eqn}
\end{equation}
In some cases, like when considering the expectation value of a single 
Wilson loop the constant is $a^2=0$, but in other cases, involving the 
correlator of two loops, $a^2$ may be positive or negative. For positive 
$a^2$ the angle $\theta$ will be a monotonous function of $\sigma$. 
For negative $a^2$, there will be an extremum for $\theta$ at some 
value $\theta_m$, where $a^2=-m^2\sin^2\theta_m$. 

In the special case when $a=0$ the solution is very simple
\begin{equation}
\sin\theta=\frac{1}{\cosh m(\pm\sigma+\sigma_i)}\,.
\label{S2-zero-a}
\end{equation}
We take the world-sheet coordinate to start at $\sigma=0$, where 
the boundary value of $\theta$ fixes $\sigma_i$ by 
$\sin\theta_i=1/\cosh m\sigma_i$. If $\sigma$ extends to 
infinity, as will be the case for the single loop, the variable 
$\theta$ will reach the north or south pole of the sphere, 
depending on the sign choice.

The action in this case is twice the kinetic term
\begin{equation}
\cS_{S^5}=2\cS_{S^5}^{kinetic}
=\frac{L^2}{4\pi\alpha'}\int d\tau\,d\sigma\,2m^2\sin^2\theta
=\frac{T\sqrt\lambda}{2\pi}\int d\theta\,|\theta'|
=\frac{T\sqrt{\lambda}}{2\pi}m|\cos\theta_f-\cos \theta_i|\,.
\label{S2-action-zero-a}
\end{equation}

Here we discussed two solutions covering the northern or southern 
hemisphere, but it is also possible to cover the sphere more times. 
Those extra wrappings are unstable world-sheet instantons, wrapping 
the sphere at a fixed point inside $AdS_5$. They can occur anywhere 
on the world-sheet, but in our symmetric ansatz they can be only at 
symmetric points. Clearly those will never give the dominant contribution 
to the action.

For $a^2>0$ it is still easy to integrate (\ref{S2-eqn}) in terms of 
elliptic integrals of the first kind with argument $\theta$ and modulus 
$im/a$
\begin{equation}
\sigma+\sigma_i
=\pm\frac{1}{a}F\left(\theta\bigg|\,i\frac{m}{a}\right)\,.
\label{S2-solution-real-a}
\end{equation}
As we will see in section~\ref{AdS5-section}, 
when $a\neq0$ the surface describes the correlator 
of two Wilson loops. Let us take $\theta_i$ and $\theta_f$ to be 
the boundary values of $\theta$ on the two loops. If the 
surface starts at $\sigma=0$ on the first loop, $\sigma_i$ is 
fixed by (\ref{S2-solution-real-a}) with $\theta=\theta_i$. 
The range of the $\sigma$ variable is given by
\begin{equation}
\delta\sigma
=\frac{1}{a}\left|
F\left(\theta_f\bigg|\,i\frac{m}{a}\right)
-F\left(\theta_i\bigg|\,i\frac{m}{a}\right)\right|\,.
\label{S2-delta-sigma-real-a}
\end{equation}
This will have to agree with the $AdS_5$ part of the ansatz.

We can also write the action in terms of elliptic integrals of the first 
and second kind. The kinetic part is
\begin{equation}
\begin{aligned}
2\cS_{S^5}^{kinetic}
=&\frac{\sqrt\lambda}{4\pi}\int d\tau\,d\sigma\,
2m^2\sin^2\theta
=\frac{T\sqrt\lambda}{2\pi}\int d\theta\,
\frac{m^2\sin^2\theta}{|\theta'|}
\\=&
\pm\frac{T\sqrt\lambda}{2\pi}a
\left[E\left(\theta\bigg|\,i\frac{m}{a}\right)
-F\left(\theta\bigg|\,i\frac{m}{a}\right)
\right]_{\theta_i}^{\theta_f}\,.
\end{aligned}
\end{equation}

The above solutions are technically also valid for $a^2<0$, but 
there are other expressions for the above elliptic 
integrals that are more suitable for this case
\begin{equation}
\sigma+\sigma_i
=\pm\frac{1}{b}F\left(\arccos\frac{\cos\theta}{\cos\theta_m}
\bigg|\,i\cot\theta_m\right)\,,
\label{S2-solution-imaginary-a}
\end{equation}
where we used $b^2=-a^2$, and $\sin\theta_m=b/m$.

This solution has a turning points at $\theta=\theta_m$ and 
$\theta=\pi-\theta_m$ 
where $\sigma+\sigma_i$ will be equal to some integral multiple 
of the complete elliptic integral. This turning point may, or may not, be 
on the world-sheet\footnote{There are also solutions with more than 
one turning point, which combine more than two branches of the solution, 
but their action will be larger than the others.}. 
To describe the solution with the turning point we 
consider the two branches with positive and negative signs in 
(\ref{S2-solution-imaginary-a}). This will have the turning point at 
$\sigma+\sigma_i=0$. If we take the solution with only one branch, 
it will not have a turning point along the world-sheet. In either case the 
value of $\sigma_i$ is fixed by plugging in the boundary value 
$\theta_i$.

The full range of $\sigma$ is now given by
\begin{equation}
\delta\sigma
=\frac{1}{b}\left|
F\left(\arccos\frac{\cos\theta_f}{\cos\theta_m}
\bigg|\,i\cot\theta_m\right)
\pm F\left(\arccos\frac{\cos\theta_i}{\cos\theta_m}
\bigg|\,i\cot\theta_m\right)\right|\,.
\label{S2-delta-sigma-imaginary-a}
\end{equation}
In the case with a turning point we have to take the positive sign, to 
add the contribution from both branches. 
The negative sign is taken when there is no turning point 
along the world-sheet.

Finally, the kinetic part of the action is
\begin{equation}
2\cS_{S^5}^{kinetic}
=\frac{T\sqrt\lambda}{2\pi}b\left[
E\left(\arccos\frac{\cos\theta_f}{\cos\theta_m}\bigg|\,
i\cot\theta_m\right)
\pm E\left(\arccos\frac{\cos\theta_i}{\cos\theta_m}\bigg|\,
i\cot\theta_m\right)
\right]\,,
\label{S2-action-imaginary-a}
\end{equation}
and the sign is chosen as in (\ref{S2-delta-sigma-imaginary-a})

A special case of the above solution occurs when 
$\theta_i=\theta_f=\pi/2$ (and $a^2<0$). Then the expression for 
the range of $\sigma$ (\ref{S2-delta-sigma-imaginary-a}) involves complete 
elliptic integrals. While one would expect $\theta_m<\pi/2$, the 
equations of motion are also solved by a constant $\theta=\pi/2$. 

This constant solution is unstable, but as we will see in 
section~\ref{AdS3xS3-section}, this case is realized when combined 
with the $AdS_5$ part. The maximal values of the complete elliptic 
integrals is for vanishing modulus, where $K(0)=\pi/2$, so 
the above ansatz allows only $\delta\sigma<\pi/m$. For a longer 
world-sheet we need to take the constant solution, which puts no 
restriction on $\delta\sigma$. The kinetic part of the action in this 
case is
\begin{equation}
\cS_{S^5}^{kinetic}=
\frac{T\sqrt\lambda}{2\pi}\,m^2\delta\sigma\,.
\label{S2-action-unstable}
\end{equation}

\subsection{$S^3$ ansatz}
\label{S3-section}

We may also consider the case with $\pi_2\neq0$ with very small 
modification. In addition to the two angles that were already turned on, 
$\theta$ and $\varphi_1$, this ansatz will include $\varphi_2$, 
but we still assume that $\rho_3=0$, or $\psi=\pi/2$. The metric 
for the three coordinates is
\begin{equation}
ds^2=d\theta^2+\sin^2\theta d\varphi_1^2+
\cos^2\theta d\varphi_2^2\,.
\end{equation}
So we are studying periodic motion on $S^3$.

In terms of $\rho_1=\sin\theta=\sqrt{1-\rho_2^2}$, the first 
integral (\ref{S5-first-integral}) reads
\begin{equation}
\rho_1'^2=a^2-\pi_2^2+(m_1^2-a^2)\rho_1^2-m_1^2\rho_1^4\,.
\label{rho-equation}
\end{equation}
This equation will have real solutions for both positive and negative $a^2$, 
but there are some constraints. First we require $|a^2+m_1^2|\geq2|m_1\pi_2|$ 
and in addition, if $a^2<\pi_2^2$, then  $m_1^2\geq a^2$.

Again this may be integrated in terms of elliptic integrals. For 
$a^2>\pi_2^2$ we write
\begin{equation}
\sigma+\sigma_i=\pm\frac{\rho_+}{\sqrt{a^2-\pi_2^2}}
F\left(\arcsin \frac{\rho_1}{\rho_+}\bigg|\,
\frac{\rho_+}{\rho_-}\right)\,,
\end{equation}
where $\rho_\pm^2$ are the two roots of the polynomial on the right-hand 
side of (\ref{rho-equation})
\begin{equation}
\rho_\pm^2=\frac{m_1^2-a^2\pm\sqrt{(a^2+m_1^2)^2-4m_1^2\pi_2^2}}{2m_1^2},.
\end{equation}

The expression for the angle $\varphi_2$ is gotten from integrating 
$\varphi_2'=\pi_2/(1-\rho_1^2)$ and is given by an elliptic integral 
of the third kind
\begin{equation}
\varphi_2=\varphi_{2i}\pm\frac{\rho_+\pi_2}{\sqrt{a^2-\pi_2^2}}
\Pi\left(\rho_+^2,\arcsin\frac{\rho_1}{\rho_+}\bigg|\,
\frac{\rho_+}{\rho_-}\right)\,.
\end{equation}
The kinetic part of the action is
\begin{equation}
\cS_{S^5}^{kinetic}=\frac{T\sqrt\lambda}{2\pi}
\frac{\sqrt{a^2-\pi_2^2}}{\rho_+}\left[
E\left(\arcsin\frac{\rho_1}{\rho_+}\bigg|\,
\frac{\rho_+}{\rho_-}\right)
-F\left(\arcsin\frac{\rho_1}{\rho_+}\bigg|\,
\frac{\rho_+}{\rho_-}\right)\right]\,.
\end{equation}
There are other expressions that will be better suited for $a^2<\pi_2^2$.

\subsection{More complicated cases}

We have studied so far cases with motion only inside an $S^3$ subspace 
of $S^5$, and turned on only one of the rotation parameters $m_1$, while 
keeping $m_2=m_3=0$.

To study the more complicated cases it may be useful to turn to the 
description of the sphere 
\cite{Babelon:1992rb,Arutyunov:2003uj}
in terms of the elliptic 
coordinates $\zeta_1$ and $\zeta_2$ that solve the equation
$\sum\frac{\rho_i^2}{\zeta-m_i^2}=0$. Assuming 
$m_1^2\leq m_2^2\leq m_3^2$, then the choice
\begin{equation}
m_1^2\leq\zeta_1\leq m_2^2\leq\zeta_2\leq m_3^2\,.
\end{equation}
will cover the range full range $\rho_i\geq0$. 
We can go back to our previous coordinates by
\begin{equation}
\begin{gathered}
\rho_1=\sqrt{\frac{(\zeta_1-m_1^2)(\zeta_2-m_1^2)}
{(m_1^2-m_2^2)(m_1^2-m_2^2)}}\,,
\qquad
\rho_2=\sqrt{\frac{(\zeta_1-m_2^2)(\zeta_2-m_2^2)}
{(m_2^2-m_1^2)(m_2^2-m_3^2)}}\,,
\\
\rho_3=\sqrt{\frac{(\zeta_1-m_3^2)(\zeta_2-m_3^2)}
{(m_3^2-m_1^2)(m_3^2-m_2^2)}}\,,
\end{gathered}
\end{equation}

In terms of those coordinates the first integrals are related to 
hyperelliptic curves, and can be solved in term of the appropriate 
integrals. Here we will not study the more complicated cases 
in any detail, as shall become clear in section~\ref{classification-section}, 
the above examples are already quite rich. We wish to make only some 
general comments about those cases.

As noted above, even if the boundary conditions fit within an $S^1$ 
equator of $S^5$, preserving an $SO(4)\times U(1)$ subgroup of 
$SO(6)$, the minimal surface will generically move off the equator 
into an $S^2$, breaking the symmetry down to $SO(3)\times U(1)$. 
If we turn on two rotations $m_1$ and $m_2$ the boundary conditions 
will be inside an $S^3$, but the classical solution will generally extend 
over an $S^4$, breaking the symmetry from $U(1)^3$ to 
$U(1)^2\times \bZ_2$.

Such solutions can be written down, but it is not clear that they will 
be minima of the action. The ansatz assumed rotational symmetry, but 
if the solution spontaneously breaks some of the symmetry, it may break 
the rotational symmetry as well.

While we do not expect this to happen in the cases we discussed above, 
it becomes totally clear that this will have to happen if we turn on 
all three of the rotation parameters $m_i$. Consider the expectation 
value of a single Wilson loop, in order to get a finite action each of the 
circles with angles $\varphi_i$ will have to shrink to zero radius, 
i.e. somewhere along the world-sheet each of the $\rho_i$ will vanish.
 
The rotationally symmetric ansatz would require this to happen on 
the same point on the world-sheet, which is impossible since 
$\sum \rho_i^2=1$. So to get a finite action the different $\rho_i$ 
have to vanish at different positions, and the rotational symmetry 
will be broken. With two rotations it's possible to have two of the 
$\rho$'s vanish simultaneously, but it is not clear that this will 
indeed be a minimum of the action. So some extra caution is 
required in addressing that case.

\section{The $AdS_5$ ansatz}
\label{AdS5-section}

We will now turn to the $AdS_5$ part of the $\sigma$-model. Again 
we consider periodic motions for the string, starting with a general ansatz 
and then specializing to several simpler cases.

The main relation between the $AdS_5$ and the $S^5$ parts of the 
ansatz comes through the 
Virasoro constraints. Since we already have the $S^5$ contribution to 
the world-sheet stress-energy tensor, the Virasoro constraints just read
\begin{equation}
\begin{aligned}
&T_{\sigma\sigma}^{AdS_5}+\frac{L^2}{8\pi\alpha'}a^2=0\,,
\nonumber\\
&T_{\sigma\tau}^{AdS_5}
+\frac{L^2}{4\pi\alpha'}\sum_i\pi_i m_i=0\,.
\end{aligned}
\end{equation}

As in the case of the sphere, a simple description of the system 
is by taking Euclidean $AdS_5$ as a hypersurface in flat six-dimensional 
Minkowski space. It is given by the hyperboloid
\begin{equation}
-Y_0^2+Y_1^2+Y_2^2+Y_3^2+Y_4^2+Y_5^2=-L^2\,.
\end{equation}

Now let us define the coordinates $r_0$, $r_1$, $r_2$, $v$, $\phi_1$ 
and $\phi_2$ by
\begin{equation}
\begin{aligned}
&Y_0=Lr_0\cosh v\,,\qquad
&Y_5=Lr_0\sinh v\,,
\\
&Y_1=Lr_1\cos\phi_1\,,\qquad
&Y_2=Lr_1\sin\phi_1\,,
\\
&Y_3=Lr_2\cos\phi_2\,,\qquad
&Y_4=Lr_2\sin\phi_2\,.
\end{aligned}
\end{equation}
Those coordinates satisfy the constraint $-r_0^2+r_1^2+r_2^2=-1$, and 
the metric of the embedding flat Minkowski space is
\begin{equation}
ds^2=L^2\left(-dr_0^2+r_0^2dv^2
+dr_1^2+r_1^2d\phi_1^2
+dr_2^2+r_2^2d\phi_2^2\right)\,.
\end{equation}

In some of the specific examples we study below we will employ 
Poincar\'e coordinates. We replace $r_0$, $r_1$, $r_2$ and $v$ with 
$\hat y$, $\hat r_1$ and $\hat r_2$ by the relations
\begin{equation}
\label{poincare-coords}
r_0=\frac{\sqrt{\hat y^2+\hat r_1^2+\hat r_2^2}}{\hat y}\,,\qquad
r_1=\frac{\hat r_1}{\hat y}\,,\qquad
r_2=\frac{\hat r_2}{\hat y}\,,\qquad
v=\ln\sqrt{\hat y^2+\hat r_1^2+\hat r_2^2}\,.
\end{equation}
In the new coordinates the metric reads
\begin{equation}
\label{poincare-metric}
ds^2=\frac{L^2}{\hat y^2}\left(d\hat y^2+d\hat r_1^2+\hat r_1^2d\phi_1^2
+d\hat r_2^2+\hat r_2^2d\phi_2^2\right )\,.
\end{equation}

Going back to the embedding coordinates, 
we consider the following ansatz, which is consistent with the equations 
of motion
\begin{equation}
r_i=r_i(\sigma)\,,\qquad
v=v(\sigma)\,,\qquad
\phi_1=k_1\tau+\alpha_1(\sigma)\,,\qquad
\phi_2=k_2\tau+\alpha_2(\sigma)\,,
\label{ads-ansatz}
\end{equation}
where $k_1$ and $k_2$ are arbitrary integers (in this case, when the 
world-sheet is compact, also the parameters $m_i$ of the $S^5$ 
ansatz have to be integers). One could consider also some $\tau$ 
dependence for $v$, but we will not include that. The action now is
\begin{equation}
\begin{aligned}
\cS_{AdS_5}=\frac{L^2}{4\pi\alpha'}\int d\sigma\,d\tau
&\big[-r_0'^2+r_1'^2+r_2'^2+r_0^2v'^2+r_1^2\alpha_1'^2
+r_2^2\alpha_2'^2+r_1^2k_1^2+r_2^2k_2^2
\\&\quad
+\Lambda\left(-r_0^2+r_1^2+r_2^2+1\right)\big]\,.
\end{aligned}
\end{equation}

$v$, $\alpha_1$ and $\alpha_2$ are cyclic, so we can express them in 
terms of the conserved momenta
\begin{equation}
v'=\frac{p_0}{r_0^2}\,,\qquad
\alpha_1'=\frac{p_1}{r_1^2}\,,\qquad
\alpha_2'=\frac{p_2}{r_2^2}\,.
\end{equation}
The equations of motion for $r_0$, $r_1$ and $r_2$ are
\begin{equation}
\begin{aligned}
r_0''&=\Lambda r_0-\frac{p_0^2}{r_0^3}\,,\\
r_1''&=(k_1^2+\Lambda)r_1+\frac{p_1^2}{r_1^3}\,,\\
r_2''&=(k_2^2+\Lambda)r_2+\frac{p_2^2}{r_2^3}\,,\\
\end{aligned}
\end{equation}

It is simple to find the first integral of motion, it's the diagonal 
component of the $AdS_5$ contribution to the stress-energy tensor,
which one can get by multiplying each of the 
above equations by the appropriate $r_i'$, summing them and using that 
$(-r_0^2+r_1^2+r_2^2)'=0$. The result is
\begin{equation}
-r_0'^2+r_1'^2+r_2'^2+\frac{p_0^2}{r_0^2}+\frac{p_1^2}{r_1^2}
+\frac{p_2^2}{r_2^2}
-r_1^2k_1^2-r_2^2k_2^2+a^2=0\,.
\end{equation}
The integration constant $a^2$ has to be the same as on the $S^5$ part of 
the action, so together the Virasoro constraint is satisfied. Using this we 
can again replace the potential terms by the kinetic ones to find for the 
classical action
\begin{equation}
\cS_{AdS_5}=\frac{\sqrt\lambda}{4\pi}\int d\sigma\,d\tau
\left[2(r_1^2k_1^2+r_2^2k_2^2)-a^2\right]
=2S_{AdS_5}^{kinetic}-
\frac{\sqrt\lambda}{4\pi}a^2 \delta\sigma T\,.
\label{AdS-classical-action}
\end{equation}
When combining this with the contribution of the sphere 
(\ref{S5-kinetic}) the $a^2$ terms will cancel each other, leaving us 
with twice the sum of the kinetic actions. This action will be divergent, 
due to a missing boundary term \cite{Drukker:1999zq}. After 
removing the divergence one finds a finite action that is negative 
(or zero). This is in contrast to the $S^5$ case, where the action was 
positive.

The other integrals of motion are
\begin{equation}
\begin{aligned}
I_0=&r_0^2-\sum_{i=1}^2\frac{1}{k_i^2}
\left((r_0r_i'-r_ir_0')^2+\frac{p_i^2}{r_i^2}r_0^2
-\frac{p_0^2}{r_0^2}r_i^2\right)\,,
\\
I_1=&r_1^2-\frac{1}{k_1^2}
\left((r_0r_1'-r_1r_0')^2+\frac{p_1^2}{r_1^2}r_0^2
-\frac{p_0^2}{r_0^2}r_1^2\right)
\\&\quad
+\frac{1}{k_1^2-k_2^2}
\left((r_1r_2'-r_2r_1')^2+\frac{p_2^2}{r_2^2}r_1^2
+\frac{p_1^2}{r_1^2}r_2^2\right)\,.
\end{aligned}
\end{equation}
We can define $I_2$ in a similar fashion, but it is not an independent 
integral, $-I_0+I_1+I_2=-1$. They are also related to the diagonal 
component of the stress-energy tensor by
\begin{equation}
k_1^2I_1+k_2^2I_2-p_0^2+p_1^2+p_2^2=a^2\,.
\end{equation}
For completeness let us write the off-diagonal component of the 
stress-energy tensor
\begin{equation}
T_{\sigma\tau}=\frac{L^2}{4\pi\alpha'}
\left(k_1p_1+k_2p_2\right)\,,
\end{equation}
which is also a constant.

So like in the case of the sphere, this periodic ansatz reduces the 
$\sigma$-model to an integrable system similar to the 
Neumann-Rosochatius system. For any value of the integration 
constants one can find the appropriate solution, though describing it 
may be complicated. As before, it is possible to introduce the 
elliptical coordinates and write the solutions in terms of hyperelliptic 
integrals.

We will not study the general solution, but instead focus below on 
some simple cases of planar concentric circles and parallel lines.

\subsection{$AdS_3$ ansatz: circles}
\label{circles-section}

The first example we look at is that of a surface that ends along two 
concentric circles at the boundary of $AdS_5$. 
In our coordinate system the boundary of $AdS_5$, which is a 
four-sphere, is given by $r_0\to\infty$. 
If we switch to the Poincar\'e patch the boundary will be flat $\bR^4$. 
In the latter case (\ref{poincare-metric}) 
we describe a circle on the boundary by a constant 
$\hat r_1$ and $\hat r_2=0$. In the former we take $r_2=0$ and 
the radius of the circle will be given by the value of $v$.

So to study concentric circles we will use our general ansatz with 
$r_2=\phi_2=0$. We can then eliminate $r_0$ from the equations by 
the identity $r_0^2=1+r_1^2$ and the Virasoro constraint turns into an 
equation for $r_1$
\begin{equation}
r_1'^2=-a^2-p_0^2-p_1^2+(k_1^2-a^2)r_1^2+k_1^2r_1^4
-\frac{p_1^2}{r_1^2}\,.
\end{equation}
It turns out to be 
useful to write the equation in terms of $z=1/r_1$, which goes to zero at 
the boundary. The above equation becomes 
\begin{equation}
z'^2=k^2+(k^2-a^2)z^2-(a^2+p_0^2+p_1^2)z^4-p_1^2z^6\,.
\label{z-equation}
\end{equation}
This can be solved in terms of elliptic integrals. In what 
follows we concentrate on $p_1=0$ and label $p_0=p$. Generically 
the surfaces will reach the boundary twice, and will correspond to 
the correlator of two Wilson loops.

\begin{figure}[ht]
\centerline{\hbox{\epsffile{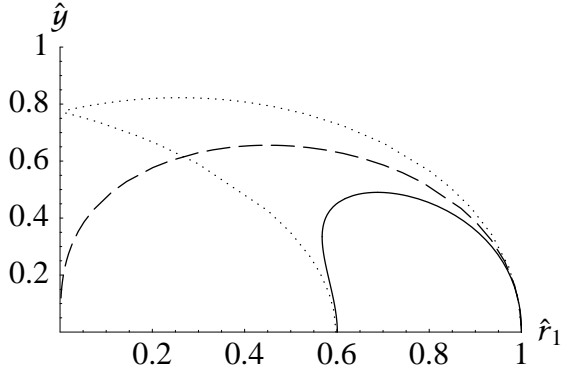}}}
\centerline{\parbox{12cm}{\caption{
\label{AdS3-figure}
The value of $\hat y$ as a function of $\hat r_1$ for three solutions to 
the $AdS_3$ ansatz. For two circles with radii $R_i=0.6$ and $R_f=1$ 
and opposite orientation (solid line), 
the solution has $a^2+p^2>0$. If the circles have the same orientation, 
the surface has to cross $\hat r_1=0$ which is given by the expressions with 
$a^2+p^2<0$ (dotted line). For $a^2+p^2=0$ (dashed line) the surface describes 
the correlator of a circle and a local operator at $\hat r_1=0$.
}}}
\end{figure}

For $a^2+p^2>0$ the equation has a turning point, a maximal value 
of $z$ before the surface goes back to the boundary. For $a^2+p^2<0$ 
(note that $a^2$ could be negative) we will have to analytically continue 
beyond $z=\infty$ to get the second part of the string. The case of 
$a^2+p^2=0$ is very interesting, and as we will see it generally describes 
the correlator of the Wilson loop with a local operator. A special case we 
will concentrate on later is for $p=0$, where $v$ is constant along 
the world-sheet. This will describe the correlator of two coincident loops, 
or in the case of $a^2=0$, the one-point function of a Wilson loop. We 
illustrate some of those solutions in figure~\ref{AdS3-figure}.

We start with the case of $a^2+p^2>0$, where the solution can be 
written in terms of elliptic integrals with modulus $z_+/z_-$ 
and argument $\arcsin z/z_+$ where $z_\pm^2$ are the 
two roots of the polynomial on the right-hand side of 
(\ref{z-equation})
\begin{equation}
z_\pm^2=\frac{k^2-a^2\pm\sqrt{(a^2+k^2)^2+4k^2p^2}}{2(a^2+p^2)}\,,
\label{circles-nu}
\end{equation}
as
\begin{equation}
\sigma=\frac{z_+}{k}
F\left(\arcsin \frac{z}{z_+}\bigg|\,\frac{z_+}{z_-}\right)
\,.
\label{circles-sigma}
\end{equation}
Inverting this equation gives $z$, or $r_0$ and $r_1$ as a function 
of $\sigma$.

Clearly $z$ has a turning point, or a maximum, at $z_+$. Beyond 
there we have to continue on the other branch of the inverse sine function, 
until reaching the boundary again. That will give the full range of the 
world-sheet coordinate $\sigma$ by the complete elliptic integrals
\begin{equation}
\delta\sigma=\frac{2z_+}{k}
K\left(\frac{z_+}{z_-}\right)\,.
\label{two-circles-delta-sigma}
\end{equation}

Next we can integrate $v$ in terms of elliptic integrals of the first and 
third kind ($v_i$ is the initial value, at $z=0$)
\begin{equation}
\begin{aligned}
v-v_i=&p\int\frac{d\sigma}{r_0^2}
=p\int\frac{dz}{z'}\frac{z^2}{1+z^2}
\\=&
\frac{pz_+}{k}\Bigg[
F\left(\arcsin\frac{z}{z_+}\bigg|\,\frac{z_+}{z_-}\right)
-\Pi\left(-z_+^2,\arcsin\frac{z}{z_+}\bigg|\,
\frac{z_+}{z_-}\right)
\Bigg]\,.
\end{aligned}
\end{equation}
This expression again covers only half the world-sheet, the other branch is 
given by a similar expression shifted by the complete elliptic integrals
\begin{equation}
\begin{aligned}
v-v_i=&
\frac{pz_+}{k}\Bigg[
2K\left(\frac{z_+}{z_-}\right)
-2\Pi\left(-z_+^2\bigg|\,
\frac{z_+}{z_-}\right)
\\&\quad
-F\left(\arcsin\frac{z}{z_+}\bigg|\,\frac{z_+}{z_-}\right)
+\Pi\left(-z_+^2,\arcsin\frac{z}{z_+}\bigg|\,
\frac{z_+}{z_-}\right)
\Bigg]\,.
\end{aligned}
\end{equation}
The surface started on the first branch at $v=v_i$. On the second branch 
$v$ reaches the final value $v_f$, so the total change is
\begin{equation}
\delta v=v_f-v_i
=\frac{2pz_+}{k}\Bigg[
K\left(\frac{z_+}{z_-}\right)
-\Pi\left(-z_+^2\bigg|\,
\frac{z_+}{z_-}\right)
\Bigg]\,.
\label{circles-delta-v}
\end{equation}

Recall that in the Poincar\'e patch (\ref{poincare-metric}) the boundary of 
$AdS_5$ is at $z=0$, and there the radius $\hat r_1=\exp v$. When 
considering 
the correlator of two concentric circles the ratio of their radii is then 
given by $\exp\delta v$. We will use this when studying specific 
examples below.

Finally we can evaluate the action. To perform the integral we have to 
take care to include the two branches, and regularize the divergence near 
the boundary by the cutoff $z_0$
\begin{equation}
\begin{aligned}
2\cS_{AdS_5}^{kinetic}
=&\frac{\sqrt\lambda}{4\pi}
\int d\tau\,d\sigma\,2k^2r_1^2
=\frac{T\sqrt\lambda}{2\pi}\,k^2\int \frac{dz}{z^2z'}
\\=&-\frac{T\sqrt\lambda}{2\pi}\,2k\Bigg[
\frac{1}{z}\sqrt{(1-\frac{z^2}{z_+^2})
\left(1-\frac{z}{z_-^2}\right)}
\\&\hskip2cm{}
-\frac{1}{z_+}F\left(\arcsin\frac{z}{z_+}\bigg|\,\frac{z_+}{z_-}
\right)+\frac{1}{z_+}E\left(\arcsin\frac{z}{z_+}\bigg|\,\frac{z_+}
{z_-}\right)
\Bigg]_{z_0}^{z_+}
\\&
=\frac{T\sqrt\lambda}{2\pi}\,2k
\Bigg[\frac{1}{z_0}
-\frac{1}{z_+}E\left(\frac{z_+}{z_-}\right)
+\frac{1}{z_+}K\left(\frac{z_+}{z_-}\right)\Bigg]\,.
\label{two-circles-action}
\end{aligned}
\end{equation}
The divergent term is proportional to the circumference of the circle, 
and is canceled by a boundary term. Thus the kinetic part of the action 
will be given just by the complete elliptic integrals. Note that in this case 
of the circles the range of the $\tau$ variable is $T=2\pi$, but we 
have chosen to leave the explicit dependence on $T$ in the expressions.

Next, we will consider the $a^2+p^2=0$ case (also studied in 
\cite{Zarembo:2002ph}), and argue that it describes
a two-point function of a Wilson loop and a local operator. We define
$b^2=-a^2$, so we have $b=\pm p$ and the solution for $z$ is simply
\begin{equation}
z=\frac{k}{\sqrt{k^2+p^2}}
\sinh\left(\sqrt{k^2+p^2}\,\sigma\right)\,,
\end{equation}
and for $v$ we get
\begin{equation}
v=p\sigma
-\hbox{arctanh}\,\left(\frac{p}{\sqrt{k^2+p^2}}
\tanh\sqrt{k^2+p^2}\,\sigma\right)\,.
\end{equation}
Both expressions diverge as $\sigma\to\infty$. To understand the geometry 
it's useful to switch to the Poincar\'e patch (\ref{poincare-metric}) 
where
\begin{equation}
\hat r_1=\frac{e^v}{\sqrt{1+z^2}}\,,\qquad
\hat y=\frac{ze^v}{\sqrt{1+z^2}}\,.
\end{equation}
If we choose the solution with negative $p$, we find that $v\to-\infty$, 
so the surface gets back to 
the boundary at $\hat r_1=0$, so instead of calculating the correlator of 
two Wilson loops, this describes the two-point function of a Wilson loop 
and a local operator at $\hat r_1=0$. The case with positive $p$, or 
$v\to\infty$ gives 
a similar surface, for a local operator inserted at infinite $\hat r_1$.

The action in this case is
\begin{equation}
2\cS_{AdS_5}^{kinetic}
=-\frac{T\sqrt\lambda}{2\pi}\sqrt{k^2+p^2}\,
\coth\sqrt{k^2+p^2}\,\sigma
\bigg|_{\sigma_{min}}^{\sigma_{max}}\,.
\end{equation}
$\sigma_{min}$ and $\sigma_{max}$ are chosen such that 
$\hat y=\hat y_0$, some cutoff. Near $\sigma=0$ there is the usual 
linear divergence which will cancel against a boundary term giving
\begin{equation}
2\cS_{AdS_5}^{kinetic}
=-\frac{T\sqrt\lambda}{2\pi}\sqrt{k^2+p^2}\,.
\end{equation}

There is another potential divergence at $\sigma\to\infty$. The full 
$AdS_5$ action (\ref{AdS-classical-action}) includes a term proportional 
to the area of the world-sheet, which will diverge logarithmically 
with $\hat y_0$. In all the cases we study this term will cancel against 
an equal contribution from the $\sigma$-model on the sphere. But it 
is possible that there are cases when those terms will not cancel exactly, 
then this logarithmic divergence will capture the anomalous dimension 
of the local operator.

For $a^2+p^2<0$, equation (\ref{z-equation}) again doesn't have a 
turning point, so the surface reaches $z=\infty$. Beyond that point we 
should continue the solution to another branch, until it comes back to 
the boundary. It again describes the correlator of two circles, but since 
the surface crosses itself at infinite $z$ (or $r_1=0$), the orientation 
of the circles will be the opposite of the previous examples. Now the 
two circles are oriented in the same direction, so the surface that connects 
them has to cross itself to preserve this orientation.

In this case both $z_+$ and $z_-$ defined in (\ref{circles-nu}) are 
imaginary, therefore it is useful to define $b^2=-a^2$ and use
\begin{equation}
\tilde z_\pm^2
=\frac{b^2+k^2\pm\sqrt{(b^2-k^2)^2+4k^2p^2}}{2(b^2-p^2)}
=-z_\pm^2\,.
\label{circles-imaginary-a-nu}
\end{equation}
Then the solution is conveniently written in terms of elliptic integrals with 
the complementary modulus $\sqrt{1-\tilde z_+^2/\tilde z_-^2}$ 
as
\begin{equation}
\sigma=\frac{\tilde z_+}{k}
F\left(\arctan\frac{z}{\tilde z_+}\bigg|\,
\sqrt{1-\frac{\tilde z_+^2}{\tilde z_-^2}}
\right)\,.
\end{equation}
Indeed we see that $z$ can extend to infinity, which corresponds 
to $r_1=0$, beyond which we have to analytically continue $\sigma$. 
The solution will reach the boundary $z=0$ again at 
$\arctan z/\tilde z_+=\pi$, so the full range of $\sigma$ is 
twice the complete elliptic integral
\begin{equation}
\delta\sigma=\frac{2\tilde z_+}{k}
K\left(\sqrt{1-\frac{\tilde z_+^2}{\tilde z_-^2}}
\right)\,.
\label{two-circles-imaginary-a-delta-sigma}
\end{equation}
Next we integrate $v$
\begin{equation}
\begin{aligned}
v-v_i=\frac{p\tilde z_+^3}{k(\tilde z_+^2-1)}\Bigg[
&F\left(\arctan \frac{z}{\tilde z_+}\bigg|\,
\sqrt{1-\frac{\tilde z_+^2}{\tilde z_-^2}}
\right)
\\&
-\Pi\left(1-\tilde z_+^2,\,\arctan\frac{z}{\tilde z_+}\bigg|\,
\sqrt{1-\frac{\tilde z_+^2}{\tilde z_-^2}}
\right)
\Bigg]\,.
\end{aligned}
\end{equation}
As before, this covers half the world-sheet, and the other branch is 
found in the same manner.
On the second branch $v$ reaches 
the final value
\begin{equation}
\delta v=v_f-v_i
=\frac{2p\tilde z_+^3}{k(\tilde z_+^2-1)}\Bigg[
K\left(\sqrt{1-\frac{\tilde z_+^2}{\tilde z_-^2}}\right)
-\Pi\left(1-\tilde z_+^2\bigg|\,
\sqrt{1-\frac{\tilde z_+^2}{\tilde z_-^2}}\right)
\Bigg]\,.
\label{circles-imaginary-a-delta-v}
\end{equation}
Finally we evaluate the action, which after removing the standard divergence 
is
\begin{equation}
2\cS_{AdS_5}^{kinetic}
=-\frac{T\sqrt\lambda}{2\pi}\,\frac{2k}{\tilde z_+}\,
E\left(
\sqrt{1-\frac{\tilde z_+^2}{\tilde z_-^2}}\right)\,.
\label{AdS3-imaginary-a-action}
\end{equation}

\subsection{$AdS_2$ ansatz: circles}
%\label{AdS2-ansatz-section}

Let us now consider the simpler case where $v$ is a constant, or $p=0$. At the 
boundary of $AdS_5$ the value of $v$ is related to the radius of 
the circle by $R=\exp v$ (\ref{poincare-coords}), so this corresponds to 
a single circle or two coincident circles on the boundary. This ansatz 
involves only the coordinates $z$ and $\phi_1$ (or $r_0$, $r_1$ and 
$\phi_1$), which parameterize an $AdS_2$ subspace of $AdS_5$, 
hence the name.

In the Poincar\'e coordinates (\ref{poincare-metric}) both $\hat y$ and 
$\hat r_1$ will be non-zero, but they will satisfy the constraint 
\begin{equation}
\hat y^2+\hat r_1^2=R^2\,,
\end{equation}
with $R$ clearly the radius of the circle on the boundary.

For $a^2>0$, the solution is the same as the more general case 
(\ref{circles-sigma}), with the replacement (\ref{circles-nu}) of 
$z_+=a/k$ and $z_-=i$ 
\begin{equation}
\sigma=\frac{1}{a}
F\left(\arcsin\frac{az}{k}\bigg|\,i\frac{k}{a}\right)\,.
\label{coincident-circles-solution}
\end{equation}
The coordinate $z$ takes values between $0$, the boundary of $AdS_5$ 
and $z_+=k/a$, where it folds back on itself, reaching again the boundary 
of $AdS_5$. 
The solution therefore covers twice a region of a Poincar\'e disk, delimited
by the boundary and a finite radius. Between the two points where the surface 
reaches the boundary, the 
world-sheet coordinate $a\sigma$ will extend from $0$ to twice 
the complete elliptic integral. Thus the range of the world-sheet 
coordinate $\sigma$ is given by
\begin{equation}
\delta\sigma=\frac{2}{a}
K\left(i\frac{k}{a}\right)\,.
\label{coincident-circles-delta-sigma}
\end{equation}

We can evaluate the action, where after accounting for the two branches, and 
removing the divergence we find
\begin{equation}
2\cS_{AdS_5}^{kinetic}=
-\frac{T\sqrt\lambda}{2\pi}\,2a\Bigg[
E\left(i\frac{k}{a}\right)
-K\left(i\frac{k}{a}\right)
\Bigg]\,.
\label{coincident-circles-action}
\end{equation}

If $a=0$ the solution is even simpler
\begin{equation}
z=\sinh k\sigma\,.
\label{one-circle-solution}
\end{equation}
In the Poincar\'e coordinates (\ref{poincare-metric}) this translates to
\begin{equation}
\hat y=R\tanh k\sigma\,,\qquad
\hat r_1=\frac{R}{\cosh k\sigma}\,,
\end{equation}
where $R=\exp v$ is the radius of the circle on the boundary.
The action in this case is 
(after subtracting the divergence)
\begin{equation}
\cS_{AdS_5}=-\frac{T\sqrt\lambda}{2\pi}\,k\,.
\label{one-circle-action}
\end{equation}

This solution is different from the case $a^2+p^2=0$ considered above. 
Here the surface does not reach the boundary again as $z\to\infty$, 
instead we reach $\hat y=R$, where the surface can connect to itself. 
Thus this surface will describe a single Wilson loop. Since $T=2\pi$, 
the action is simply $k$ times the result for the simplest 
circular observable \cite{Berenstein:1999ij,Drukker:1999zq}.

For $a^2<0$, the $z$ coordinate again extends to infinity, or $r_1$ reaches 
zero. But now the surface will not close smoothly on itself. Instead we have 
to continue it beyond that point, until it reaches the boundary again.
Defining $b^2=-a^2$ we have the general case with the replacement 
$\tilde z_+=1$ and $\tilde z_-=k/b$
\begin{equation}
\sigma=\frac{1}{k}
F\left(\arctan z\bigg|\,\sqrt{1-\frac{b^2}{k^2}}\right)\,.
\label{coincident-circles-imaginary-a}
\end{equation}
The range of $\sigma$ is again twice the complete elliptic integral, which 
we write in two ways utilizing the symmetry of exchanging the two roots 
$\tilde z_+\leftrightarrow\tilde z_-$
\begin{equation}
\delta\sigma
=\frac{2}{k}K\left(\sqrt{1-\frac{b^2}{k^2}}\right)
=\frac{2}{b}K\left(\sqrt{1-\frac{k^2}{b^2}}\right)\,,
\end{equation}
and the action is
\begin{equation}
2\cS_{AdS_5}^{kinetic}=
-\frac{T\sqrt\lambda}{2\pi}\,2k\,
E\left(\sqrt{1-\frac{b^2}{k^2}}\right)
\,.
\label{coincident-circles-action-imaginary-a}
\end{equation}

\subsection{$AdS_3$ ansatz: straight lines}
\label{lines-section}

We now wish to consider the case of infinite anti-parallel straight 
lines. This is a degenerate example of the planar circles, which corresponds 
to taking 
the double limit of nearly coincident circles of very large radii. 
Yet, since the anti-parallel lines are very natural observables, which 
capture the potential between external charged particles, we 
derive the results in detail.

Consider two lines extended in the $x_1$ direction separated in 
the $x_2$ direction at the boundary of the Poincar\'e patch. This 
naturally leads to the ansatz (in what follows we omit the hats to avoid 
clutter), 
\begin{equation}
x_1=\tau\,,\qquad
y= y(\sigma)\,,\qquad
x_2= x_2(\sigma)\,,
\end{equation}
which produces the following $AdS_5$ action
\begin{equation}
\cS_{AdS_5}=\frac{L^2}{4\pi\alpha'}\int d\sigma\, d\tau\, 
\frac{1+y'^2+x_2'^2}{y^2}\,.
\end{equation}
The coordinate $x_2$ is cyclic, with a conserved momentum 
$p=x_2'/y^2$.

The equation of motion for $y$
\begin{equation}
\frac{yy''-y'^2+1}{y^3}+p^2y=0\,,
\end{equation}
is integrated once to yield
\begin{equation}
\frac{y'^2-1}{y^2}+p^2y^2=-a^2\,.
\label{y-equation}
\end{equation}
Note that as before, the left hand side is proportional to the $AdS_5$ 
contribution to the diagonal part of the stress-energy tensor, 
therefore the constant on the right hand side has to be equal to the 
integration constant on the sphere. And again we express the on-shell action 
in terms of the kinetic term
\begin{equation}
\cS_{AdS_5}=\frac{\sqrt\lambda}{4\pi}\int d\sigma\,d\tau\,
\left(\frac{2}{y^2}-a^2\right)
=2S_{AdS_5}^{kinetic}
-\frac{\sqrt\lambda}{4\pi}\int d\sigma\,d\tau\,a^2
\end{equation}

Comparing the equation for $y$ (\ref{y-equation})
to the equation for $z$ above (\ref{z-equation}) we 
see that this is a degenerate case when we replace
\begin{equation}
z\to ky\,,\qquad
p\to\frac{p}{k}\,,
\end{equation}
and take the limit of $k\to0$. So all the formulae below will follow 
from the previous ones in this limit.

Unless $p^2=0$ and $a^2\leq 0$, $y$ will have a turning point at
\begin{equation}
y_+^2=\frac{-a^2+\sqrt{a^4+4p^2}}{2p^2}\,.
\label{lines-y+}
\end{equation}
The equation can be solved in terms of an elliptic integral of the first kind
with modulus $ipy_+^2=y_+/y_-$, where $y_-$ is the other root of 
(\ref{y-equation})
\begin{equation}
\sigma=y_+
F\left(\arcsin\frac{y}{y_+}\bigg| \,ipy_+^2\right)\,,
\label{lines-solution}
\end{equation}
Varying $y$ between zero and $y_+$ 
corresponds to $\sigma/y_+$ going from $0$ to the complete elliptic 
integral. Unless we study the single straight line, this range covers 
only half the world-sheet, so we have to take care to multiply some 
quantities by two to fix that. Thus
\begin{equation}
\delta\sigma=2y_+
K\left(i py_+^2\right)\,.
\label{lines-delta-sigma}
\end{equation}

$x_2$ in turn can be written in terms of elliptic integrals of the first and 
second kind
\begin{equation}
x_2=p\int d\sigma\,y^2
=\frac{1}{py_+}
\left[E\left(\arcsin\frac{y}{y_+}\bigg| \,ipy_+^2\right)-
F\left(\arcsin\frac{y_+}{y}\bigg| \,ipy_+^2\right)\right]\,.
\label{x2}
\end{equation}
The distance between the lines is given by the complete integrals
\begin{equation}
R=\frac{2}{py_+}\left[E\left(ipy_+^2\right)
-K\left(ipy_+^2\right)\right]\,.
\label{distance-lines}
\end{equation}
The constant $a$ is 
determined from the solution of the $S^5$ equation, leaving this 
expression for $R$ to fix the integration constant $p$. Scale invariance 
of the theory means that there is a simple scaling law, $p^{1/2}R$ 
depends only on the ratio $p/a^2$.

We can calculate the action, by the same integrals as in the case of the 
circle (\ref{coincident-circles-action}), accounting for the two halves of the 
world-sheet and removing the divergence we find
\begin{equation}
2\cS_{AdS_5}^{kinetic}
=-\frac{T\sqrt\lambda}{2\pi}\,\frac{2}{y_+}\left[
E\left(ipy_+^2\right)
-K\left(ipy_+^2\right)\right]\,,
\label{lines-action}
\end{equation}
where $T$ is the length of the lines

\subsection{$AdS_2$ ansatz: straight lines}

A simple case is when $p=0$, which means the solution has no 
dependence on the $x_2$ direction, or in other words $R=0$ 
(or $R\to\infty$). Like in the case of the circles, this is the 
$AdS_2$ degeneration of the $AdS_3$ ansatz. This solution will 
describe two coincident lines or a single line.

For $a^2>0$ the solution is
\begin{equation}
y(\sigma)=\frac{1}{a}\sin a\sigma\,,
\label{coincident-lines-solution}
\end{equation}
and the surface reaches the boundary twice, at $\sigma=0$ and 
at $\sigma=\pi/a$. 
The kinetic part of the action is
\begin{equation}
\cS_{AdS_5}^{kinetic}
=\frac{\sqrt\lambda}{2\pi}\int d\sigma\,d\tau
\frac{1}{y^2}
=\frac{T\sqrt\lambda}{2\pi}
\frac{2}{y_0}\,.
\end{equation}
So after removing the divergence the action vanishes.

In the special case when $a^2=0$ the solution is even simpler, 
$y(\sigma)=\sigma$, and it describes a single straight line. The 
action again will vanish.

Finally, for $a^2<0$ the solution is better written  in terms of 
$b^2=-a^2$ as
\begin{equation}
y(\sigma)=\frac{1}{b}\sinh b\sigma\,.
\end{equation}
This solution extends to infinite $y$, but like in the case of the circle 
with negative $a^2$, it carries momentum at infinity, and perhaps should 
be connected to another solution there describing the correlator of two 
lines with the same orientation. The range of $\sigma$ now 
diverges, but since we have not found corresponding solutions to the $S^5$ 
ansatz, so we cannot realize this example.

\subsection{More complicated cases}

We have discussed certain solutions that fit within $AdS_2$ and 
$AdS_3$ subspaces 
of $AdS_5$, or Wilson loop operators that fit within a plane (or an $S^2$) 
on the boundary. But our general ansatz allows much more general 
solutions.

First, still within the $AdS_3$ ansatz it is possible to consider the case 
where $p_1\neq0$. The solution of equation (\ref{z-equation}) is still 
an elliptic integral. So the expressions will be similar to the above, only 
somewhat more involved. The real difference comes because now the 
off-diagonal Virasoro constraint, which includes $k_1p_1$, will not be 
satisfied within $AdS_5$ alone. To fix that we have to take $m_1\pi_1$ 
in the $S^5$ ansatz non-zero too.

The effect of this is to include some extra phase shift along the 
world-sheet. If one considers, say, the two circles, one boundary will 
be given by $\phi_{1i}=k_1\tau$, and the other by 
$\phi_{1f}=\alpha_1(\delta\sigma)+k_1\tau$. The relative phase 
will be non-zero if $p_1\neq0$, and will be meaningful only if there 
is some rotation (say $m_1$) on $S^5$. Therefore turning on $p_1$ 
corresponds exactly to that case with a relative phase between the lines. 
We will not study this case in detail.

This generalization is still within the $AdS_3$ subspace, but our ansatz 
allows much more general solutions, mainly turning on $k_2$. That 
will correspond to a Wilson loop that wraps two circles on orthogonal 
planes on the boundary. Again we can consider the one-point function of 
this operator, or the correlator of two. Another case is when we take 
the radius of one of the circles to infinity, which will give a helix. 
A solution corresponding to the correlator of two helices was already 
presented in \cite{Mikhailov:2002ya}.

We have not studied those cases in detail, and in particular have not checked 
whether the periodic ansatz used here will always give the true minimum 
of the action. As mentioned above, in the $S^5$ case, there are reasons to 
believe that the symmetric solution will not always give the minimum of 
the action, so one should take care in studying these examples.

\section{Classification of solutions}
\label{classification-section}

After studying those general solutions on the sphere and in $AdS_5$ we 
put them together here, pointing out special features that arise in the 
different examples. A lot of the examples were studied over the past 
years, we try to collect the known facts about those cases, and discuss 
some new solutions.

\subsection{$AdS_2$ subspace}
\label{AdS2-section}

We start with a very familiar example, where the Wilson loop couples only 
to one of the scalars, leading to a trivial $S^5$ ansatz, and also the spatial 
part is the simplest, either a single line, or a circle.

The straight line is given by (\ref{line-loop}) and 
(\ref{S5-loop}) with $\theta=\pi/2$ and $m_1=0$. 
The solution for the $S^5$ part is trivial leading to $a^2=0$, 
and in $AdS_5$ the solution is simply
\begin{equation}
x_1=\tau\,,\qquad
y=\sigma\,,
\label{line-solution}
\end{equation}
This Wilson loop preserves half of the supersymmetry of the 
theory, and its expectation value is trivial $\vev{W}=1$. The surface 
described by the above coordinates has a geometry of $AdS_2$.

A more interesting example is the circular Wilson loop 
(\ref{circle-loop}), with arbitrary wrapping $k$. 
The minimal surface is given by ({\ref{one-circle-solution})
\begin{equation}
\phi_1=k\tau\,,\qquad
z=\sinh k\sigma
\end{equation}
or in the other coordinate system
\begin{equation}
\phi_1=k\tau\,,\qquad
\hat r_1=\frac{R}{\cosh k\sigma}\,,\qquad
\hat y=R\tanh k\sigma\,.
\end{equation}
This surface again has the geometry of $AdS_2$, but now the action 
(\ref{one-circle-action}) 
is given by $\cS=-k\sqrt\lambda$, so the expectation value of the 
Wilson loop is
\begin{equation}
\vev{W}=\exp k\sqrt\lambda\,.
\end{equation}

This Wilson loop also preserves half the supersymmetries 
\cite{Bianchi:2002gz,Mikhailov:2002ya}, but all the supercharges 
it preserves involve the superconformal generators, so they do not 
close on the Hamiltonian, and apparently do not force the expectation 
value to be trivial.

In fact the straight line and the circle are related by a conformal 
transformation. The difference between the two was attributed 
\cite{Drukker:2000rr} to an anomaly that arises since the conformal 
transformation takes the point at infinity where the line ends, to 
a finite distance.

Quite remarkably one can reproduce this result from a perturbative 
calculation assuming only ladder/rainbow diagrams contribute (in the 
Feynman gauge) \cite{erickson}. The exact result at finite $N$ from 
this gauge theory calculation is captured by the Hermitian matrix 
model given by the following integral over all 
$N\times N$ Hermitian matrices $M$
\begin{equation}
\vev{W_{\rm ladders}}=\vev{\frac{1}{N} \Tr\exp M}
=\frac{1}{Z}\int {\cal D}M
\frac{1}{N} \Tr (\exp kM)
\exp\left(-\frac{2N}{\lambda} \Tr M^2\right)\,.
\end{equation}
The leading behavior at large~$N$, expressed in terms of the modified 
Bessel function, is easily found using Wigner's 
semi-circle law
\begin{equation}
\vev{W_{\rm ladders}}
\sim \int_{-1}^1
dx\, \sqrt{1-x^2}\,\exp\left(xk\sqrt{\lambda}\right)
=\frac{2}{k\sqrt\lambda}I_1\left(k\sqrt\lambda\right)
\sim \exp k\sqrt\lambda\,.
\end{equation}
This is indeed the leading behavior of the circular Wilson loop as 
calculated by the string in $AdS_5$.

One can do better and solve this matrix model exactly applying several 
different techniques. Using orthogonal polynomials, the full result at 
finite $N$ was given \cite{Drukker:2000rr,Staudacher:1997kn} 
in terms of a Laguerre polynomial 
$L_n^k(x)=1/n!\exp[x]x^{-k}(d/dx)^n(\exp[-x]x^{n+k})$
as
\begin{equation}
\vev{W_{\rm ladders}}
=\vev{\frac{1}{N}\Tr\exp M}
=\frac{1}{N}L_{N-1}^1\left(-\frac{k^2\lambda}{4N}\right)
\exp\left(\frac{k^2\lambda}{8N}\right)\,,
\label{laguerre}
\end{equation}
Several properties of this result, expanded at large~$N$ and $\lambda$ 
were compared to the expected behavior of semiclassical string in 
$AdS_5$.

The most extensive test of this expression was carried out in 
\cite{Drukker:2005kx} where all the non-planar corrections 
at large~$N$ and large $\lambda$ with fixed ration $k^2\lambda/N$ 
were evaluated. The result
\begin{equation}
\exp\left[-\frac{k\sqrt\lambda}{2}
\sqrt{1+\frac{k^2\lambda}{16N^2}}
-2N\hbox{arcsinh}\,\frac{k\sqrt\lambda}{4N}\right]\,,
\end{equation}
was then compared to a calculation of the Wilson loop using a D3-brane 
instead of a fundamental string. The results were in complete agreement 
capturing all quantum corrections beyond the leading planar result.

While this $AdS_2$ sector includes non-trivial Wilson loops only along 
the circle, there is an infinite family of Wilson loops one may consider. 
Here we discussed only the single trace operator of the Wilson loop 
wrapped $k$ times, but there are multi-trace Wilson loops. Some 
were calculated in \cite{Drukker:2000rr}, and the resulting 
expressions are rather complicated. Alternatively, one may consider 
Wilson loops in higher dimensional representations of the gauge 
group.

\subsection{$AdS_3\times S^1$ subspace}
\label{AdS3xS1-section}

The $AdS_2$ example has many interesting features, but in terms 
of finding the minimal surface solution, it's extremely trivial. So we 
turn now to the $AdS_3\times S^1$ subspace. 
In the previous example the Wilson loop was along a one-dimensional 
line or circle, and the resulting surface had the geometry of $AdS_2$. 
In general if the Wilson loop is planar, the resulting minimal surface will 
sit within an $AdS_3$ subspace of $AdS_5$, since the 
solution will depend only on the two coordinates on the plane and 
the radial coordinate $\hat y$.

The same is true for operators that are defined along a curve inside any 
2-sphere on the boundary of $AdS_5$. One can stereographically 
project any 2-sphere to a plane, and unless a point 
along the Wilson loop is mapped to infinity, the result of the calculation 
will not be altered.

The periodic solutions that fit 
in this sector correspond to two anti-parallel lines 
(i.e. (\ref{line-loop}) at two values of $x_2$ separated 
by $R$), or two concentric circles (i.e. (\ref{circle-loop}) 
with radii $R_i$ and $R_f$). In addition we allow the two lines to 
couple to different scalars, (i.e. (\ref{S5-loop}) with 
$\theta=\pi/2$, $m_1=0$ and $\varphi_1=\varphi_{1i}$ on one line 
and $\varphi_1=\varphi_{1f}$ on the second).

The minimal solution will fit within an $S^1\subset S^5$. That ansatz 
has $a^2>0$ which will select the $a^2+p^2>0$ case of the $AdS_3$
ansatz.

\subsubsection{Anti-parallel lines}
\label{AdS3xS1-lines-section}

Here we study the example of two infinite anti-parallel straight 
lines, extended in the $x_1$ direction and located at $x_2=\pm R/2$. 
In addition to the separation in space, we allow the lines to 
be at different points on $S^5$. So the boundary conditions 
along the first line are
\begin{equation}
x_1=\tau\,,\qquad
x_2=-\frac{R}{2}\,,\qquad
\varphi_1=\varphi_{1i}\,,
\end{equation}
and along the second line
\begin{equation}
x_1=\tau\,,\qquad
x_2=\frac{R}{2}\,,\qquad
\varphi_1=\varphi_{1f}\,.
\end{equation}

This type of Wilson loops was the one studied in the original papers of 
Rey and Yee and of Maldacena \cite{rey-wl,maldacena-wl}. 
Already in \cite{maldacena-wl} the two lines were allowed to be at 
different positions on $S^5$. As explained in section~\ref{S1-section}, 
this corresponds to turning on only the integration constant $\pi_1$ 
in (\ref{pi's}). All the $m_i$ in our general ansatz 
(\ref{S5-ansatz}) as well as $\pi_2$ and $\pi_3$ are set to 
zero and we take $\rho_1=1$ and $\rho_2=\rho_3=0$.

So only the angle $\varphi_1$ will vary along the world-sheet and is 
given by $\varphi_1=\varphi_{1i}+\pi_1\sigma$. The constant $a$ in 
the Virasoro constraint is given by $a=\pi_1$. and the range of the 
world-sheet coordinate $\sigma$ is
\begin{equation}
\label{anti-parallel-lines-delta-sigma1}
\delta\sigma=\frac{\delta\varphi_1}{a}\,,
\end{equation}
where the two lines are separated by an angle $\delta\varphi_1$.

In the general ansatz for parallel lines (section~\ref{lines-section}) 
we use the same constant $a$ and the range of $\sigma$ was given by 
the complete elliptic integral (\ref{lines-delta-sigma})
\begin{equation}
\label{anti-parallel-lines-delta-sigma2}
\delta\sigma=2y_+
K\left(ipy_+^2\right)\,.
\end{equation}
with (\ref{lines-y+})
\begin{equation}
y_+^2=\frac{-a^2+\sqrt{a^4+4p^2}}{2p^2}\,.
\end{equation}

Equating those two expressions for $\delta\sigma$ gives a relation 
between $\delta\varphi_1$ and $p/a^2$. This simple scaling is a 
consequence of conformal invariance, the distance between two 
parallel lines can be changed by a conformal transformation. 
Equation~(\ref{distance-lines})
\begin{equation}
\label{anti-parallel-lines-R}
R=\frac{2}{y_+p}\left[E\left(ipy_+^2\right)
-K\left(ipy_+^2\right)\right]\,,
\end{equation}
gives $R\sqrt{p}$ as a function of $p/a^2$, and then by the previous 
relation in terms of $\delta\varphi_1$.

The full action for this solution is given by twice the sum of the kinetic 
terms on $S^5$, which vanishes in this case, and the $AdS_5$ part 
(\ref{lines-action}). The result is
\begin{equation}
\label{actionconn}
\cS=-\frac{T\sqrt{\lambda}}{2\pi}\,\frac{2}{y_+} 
\left[E\left(ipy_+^2\right)
-K\left(ipy_+^2\right)\right]
=-\frac{T\sqrt\lambda}{2\pi}pR\,.
\end{equation}
This is a simple way of writing the action, but the right-hand-side is 
not expressed solely in 
terms of the geometric data, $R$ and $\delta \varphi_1$. To fix that 
one uses the above relations to replace for $p$, which will be $1/R^2$ 
times some function of $\delta\varphi_1$. Then the right-hand 
side will always exhibit a Coulomb-like potential.

If the two lines are 
at the same value of $\varphi_1$, the solution simplifies. We set $a=0$, and 
find \cite{rey-wl,maldacena-wl}
\begin{equation}
R=\frac{2}{\sqrt p}\left[E(i)-K(i)\right]
=\frac{(2\pi)^{3/2}}{\Gamma(1/4)^2\sqrt{p}}\,.
\label{AdS3xS1-R}
\end{equation}
Using (\ref{lines-action}) we evaluate the action, which after subtracting 
the divergence is
\begin{equation}
\cS=-\frac{\sqrt\lambda}{2\pi}(R\sqrt{p})^2\frac{T}{R}
=-\frac{4\pi^2\sqrt\lambda}{\Gamma(1/4)^4}\frac{T}{R}\,.
\end{equation}

It is worth pointing out that even in this simple case, the boundary 
conditions allow for another solution: two disconnected surfaces as in 
equation 
(\ref{line-solution}) corresponding each to a single straight line. The action 
for this solution is zero, and the action for the connected solution 
(\ref{actionconn}) is always negative, so in this case the connected solution
dominates for any value of the geometric data, $R$ and $\delta \varphi_1$.
In more general cases, different solutions will dominate in different ranges
of the parameters, as we will see in the next example.

\subsubsection{Concentric circles}
\label{concentric-circles-section}

A richer example is that of two concentric circles in a plane in $\bR^4$. 
The new feature that arises is that for certain values of the radii, the 
connected classical solution ceases to exist. This phenomenon was described 
in this context first by Gross and Ooguri \cite{Gross:1998gk}. They 
used the intuition from flat space where two circles on 
parallel planes have a connected minimal surface between them only 
if they are separated by a distance smaller than roughly $1.325$ times 
their radius.

This configuration was studied in $AdS_5$ space in 
\cite{Zarembo:1999bu,Olesen:2000ji}, where two concentric circles 
of equal or unequal radii on two parallel planes where considered. In our 
ansatz we study circles in the same plane, but those solutions are not 
really different. Two concentric circles on parallel planes define a 
2-sphere or a plane in $\bR^4$. 
Since we can relate any 2-sphere to the plane by a conformal 
transformation, those systems are equivalent. We repeat the calculation 
here, generalizing it by allowing the two circles to have different values 
for the $S^5$ angle $\varphi_1$.

Now the boundary conditions on the string along the first circle are
\begin{equation}
\phi_1=k\tau\,,\qquad
v=v_i\,,\qquad
\varphi_1=\varphi_{1i}\,,
\end{equation}
and along the second circle
\begin{equation}
\phi_1=k\tau\,,\qquad
v=v_f\,,\qquad
\varphi_1=\varphi_{1f}\,.
\end{equation}
By conformal invariance, the result ought to depend only on the ratio
of the radii, which according to (\ref{poincare-coords}) is given by
$R_f/R_i=\exp (v_f-v_i)$.

As in the case of the parallel lines, we have to relate the range of the 
world-sheet coordinate $\sigma$ in the two parts of the ansatz. From 
(\ref{S1-delta-sigma}) and (\ref{two-circles-delta-sigma}) we get the 
relation
\begin{equation}
\delta\sigma
=\frac{\delta\varphi_1}{a}
=\frac{2z_+}{k}
K\left(\frac{z_+}{z_-}\right)\,,
\end{equation}
with (\ref{circles-nu})
\begin{equation}
z_\pm^2=\frac{k^2-a^2\pm\sqrt{(a^2+k^2)^2+4k^2p^2}}{2(a^2+p^2)}\,,
\end{equation}

A second equation comes from the boundary conditions on $v$. The 
total shift in $v$ is given by
(\ref{circles-delta-v})
\begin{equation}
\delta v
=v_f-v_i
=\log\frac{R_f}{R_i}
=\frac{2pz_+}{k}\Bigg[
K\left(\frac{z_+}{z_-}\right)
-\Pi\left(-z_+^2\bigg|\,
\frac{z_+}{z_-}\right)
\Bigg]\,.
\end{equation}

The full action is again twice the sum of the kinetic terms, which 
vanishes for this $S^1$ ansatz, so it's just the $AdS_5$ contribution 
restricted to the $AdS_3$ case (\ref{two-circles-action}), with 
$T=2\pi$
\begin{equation}
\cS=-\sqrt{\lambda}\,\frac{2k}{z_+ }
\Bigg[E\left(\frac{z_+}{z_-}\right)
-K\left(\frac{z_+}{z_-}\right)\Bigg]\,.
\end{equation}

One can always construct also a disconnected solution, each of the circles 
will be the boundary of the world-sheet described in 
section~\ref{AdS2-section}. The action for the disconnected solution 
doesn't depend on the ratio of radii, it's simply twice the action there, 
$\cS=-2k\sqrt\lambda$. The Gross-Ooguri phase transition 
\cite{Gross:1998gk} will take place when the action of the connected 
solution reaches this value. In the case with $\delta \varphi_1=0$ this 
happens at $R_f/R_i\sim2.4034$.

Furthermore, for any fixed value of $\delta \varphi_1$, by plotting 
$\delta v$, one can see that it reaches a maximum, which implies
that we have connected surfaces only for a finite range of the ratio of radii. 
Beyond that, only the disconnected solution exists. For $\delta \varphi_1=0$,
this happens for $K(z_+/z_-)=2E(z_+/z_-)/(1-z_+^2/z_-^2)$ 
or roughly 
$p/k\sim0.5811$. This value corresponds to a ratio $R_f/R_i\sim 2.7245$, and 
was already obtained by Zarembo and Olesen 
\cite{Zarembo:1999bu,Olesen:2000ji}, who considered the 
correlator of two circles in parallel planes 
separated by a distance $L$. Their configuration and ours are conformally 
related by a stereographic projection.

We illustrate this in figure~\ref{AdS3xS1-figure}. A connected solution 
exists only inside the dashed curve. Inside the solid 
curve this connected solution has lower action than the disconnected one 
and will dominate. The closer the initial and final values of $R$ are, 
the range of $\delta\varphi_1$ with a connected solution will 
increase. This graph is symmetric under the inversion 
$R_f/R_i\to R_i/R_f$.

\begin{figure}[ht]
\centerline{\hbox{\epsffile{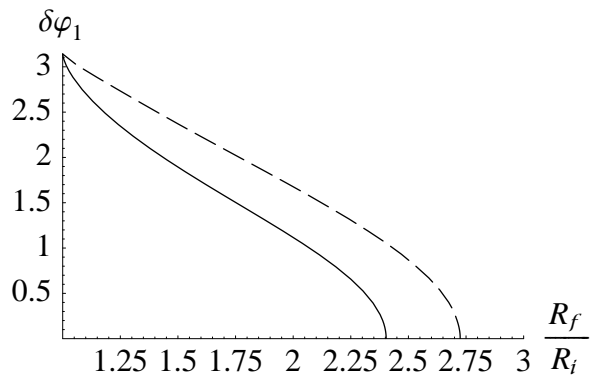}}}
\centerline{\parbox{12cm}{\caption{
\label{AdS3xS1-figure}
The allowed range of $R_f/R_i$ for two circles of radii $R_f\geq R_i$ 
as a function of the separation on the sphere, 
$\delta\varphi_1$. Connected classical solutions exist for all values 
inside the region bound by the dashed line. This connected solution 
dominates only inside the region bound by the solid line.
}}}
\end{figure}

\subsection{$AdS_2\times S^2$ subspace}
\label{AdS2xS2-section}

We consider now the case where on the boundary of $AdS_5$ the source 
is localized along a line or a circle and on the $S^5$ part the ansatz will 
involve periodic motion inside a 2-sphere. This ansatz includes single 
lines, or circles with some rotation on $S^2$, and may also include 
the correlator of two lines/circles, as long as they are separated only in 
the $S^5$ directions. This sector obviously includes the line and 
circle with no rotation reviewed in section~\ref{AdS2-section}. 
Another case already studied in section~\ref{AdS3xS1-section} is 
that of two coincident lines or circles separated on the sphere by a 
constant angle $\delta\varphi_1$.

\subsubsection{Single line}
\label{AdS2xS2-line}

We first review the case of a single straight line, already presented 
in \cite{Tseytlin:2002tr}. The Wilson loop will be the line 
(\ref{line-loop}) with periodic coupling to three of the scalars, 
(\ref{S5-loop}) with $\cos\psi=0$ and only $m_1=m$ non-zero.
The minimal surface will be given by the $S^2$ solution discussed 
in section~\ref{S2-section} in the special case of $a=0$, and the 
simplest $AdS_2$ ansatz. The solution is
\begin{equation}
x_1=\tau\,,\qquad
y=\sigma\,,\qquad
\varphi_1=m\tau\,,\qquad
\sin\theta=\frac{1}{\cosh m(\pm\sigma+\sigma_i)}\,.
\end{equation}
The world-sheet extends from $\sigma=0$ to infinity and $\sigma_i$ 
is set by the boundary value of $\theta$ as 
$\sin\theta_i=1/\cosh m\sigma_i$. The choice of sign corresponds 
to the surface covering the northern or southern hemisphere, one of which 
will be a minimum and the other an unstable saddle point (unless 
$\theta_i=\pi/2$).

The $AdS_5$ part of the solution is identical to the usual straight line 
(\ref{line-solution}), whose action vanishes (after including the boundary 
term). The action therefore comes from the $S^5$ part, and is equal to 
the area of the part of the sphere covered by the surface a certain number 
of times (\ref{S2-action-zero-a})
\begin{equation}
\cS=\frac{\sqrt\lambda}{2\pi}Tm
\left(1-|\cos\theta_i|\right)\,,
\end{equation}
where $T$ is a regulator of the length of the line.

\subsubsection {Single circle}
\label{AdS2xS2-circle-section}

Next we consider the expectation value of a Wilson loop which is 
wrapped $k$ times around a circle (\ref{S5-loop}) while wrapping 
$m$ times a parallel at angle $\theta_i$ on $S^2$.

Again the constant $a$ vanishes 
so the solution of the sphere $\sigma$-model is like in the above 
example, while for the $AdS_5$ part we use (\ref{one-circle-solution}) 
giving
\begin{equation}
\phi_1=k\tau\,,\qquad
z=\sinh k\sigma\,,\qquad
\varphi_1=m\tau\,,\qquad
\sin\theta=\frac{1}{\cosh m(\pm \sigma+\sigma_i)}\,,
\end{equation}
Again the sign corresponds to a surface extending over the northern or 
southern hemisphere and is chosen to minimize the action, and at 
$\sigma=0$ the boundary value of $\sin \theta$ is set to 
$1/\cosh m\sigma_i$. We also write the solution in terms of the 
coordinates on the Poincar\'e patch
\begin{equation}
\hat y=R\tanh k\sigma\,,\qquad
\hat r_1=\frac{R}{\cosh k\sigma}\,.
\end{equation}
Note that $\hat y^2+\hat r_1^2=R^2$.

The contribution to the action from the $AdS_5$ part 
(\ref{one-circle-action}) is $k$ times that 
of the regular circle, or $\cS_{AdS_5}=-k\sqrt\lambda$. From the 
$S^5$ part we get the area of the part of the sphere covered by the 
surface $\cS_{S^5}=m\sqrt\lambda(1-|\cos\theta_i|)$. Together 
the action is
\begin{equation}
\cS=\left(-k+m-m|\cos\theta_i|\right)\sqrt\lambda\,.
\end{equation}
Thus the expectation value of the Wilson loop at strong coupling is given 
by
\begin{equation}
\vev{W}\sim \exp\left[
(k-m+m|\cos\theta_i|)\sqrt\lambda\right]\,,
\end{equation}

Note here that in the special case when $m=k$ this reduces to
\begin{equation}
\vev{W}\sim \exp\left[
k\cos\theta_i\sqrt\lambda\right]\,,
\label{ads-vev}
\end{equation}
which will be studied in detail in \cite{we}.

In the specific case when $\theta_i=\pi/2$, this Wilson loop preserves 
$1/4$ of the supersymmetries of the theory, fitting a general ansatz 
for supersymmetric Wilson loops found by Zarembo 
\cite{Zarembo:2002an}. The action for the string solution in this 
case vanishes, giving the Wilson loop expectation value unity.
It was subsequently proven that this Wilson loop is equal to one 
to all orders in perturbation theory 
\cite{Guralnik:2003di,Guralnik:2004yc}.

If $\theta_i=0$ the surface will stay at the north pole, so the 
$S^5$ ansatz is trivial, this is just the $AdS_2$ solution discussed in 
section~\ref{AdS2-section}.

\subsubsection{Coincident circles}
\label{coincident-circles-section}

The most complicated example in this sector is that of two coincident 
circular loops. The two circles may be oriented in the same way, or 
in opposite directions, and as above we allow a periodic coupling to 
the scalars.

The case of oppositely oriented circles is rather subtle. If we considered 
coincident lines the potential between them would diverge. One way 
to see that is to consider the solution with the lines separated, where 
they will exhibit a Coulomb potential, which will diverge when 
they coincide. The surface describing them will get closer and closer 
to the boundary and in the limit it becomes singular.

Another way of seeing this is to take the solution for coincident lines 
(\ref{coincident-lines-solution}) with $a^2>0$. The range of 
$\sigma$ is then $\pi/a$ and has to be equal to that on the sphere, 
which for the $S^2$ ansatz (\ref{S2-delta-sigma-real-a}) is
\begin{equation}
\delta\sigma=\frac{\pi}{a}
=\frac{1}{a}\left|
F\left(\theta_f\bigg|\,i\frac{m}{a}\right)
-F\left(\theta_i\bigg|\,i\frac{m}{a}\right)\right|\,.
\end{equation}
The right-hand satisfies the following inequalities
\begin{equation}
\left|F\left(\theta_f\bigg|\,i\frac{m}{a}\right)
-F\left(\theta_i\bigg|\,i\frac{m}{a}\right)\right|
\leq2K\left(\,i\frac{m}{a}\right)
\leq\pi\,,
\end{equation}
where $\theta_i$ and $\theta_f$ are the boundary values of $\theta$ 
on the two circles. 
Hence the equation for $\delta\sigma$ can only be satisfied for 
$\theta_i=0$ and $\theta_f=\pi$, where this solution is not the 
dominant one.

The same is true if we take the $S^1$ ansatz (\ref{S1-delta-sigma}), 
where the range of $\sigma$ is $\delta\sigma=\delta\varphi_1/a$, 
and again there will be no solution for $\delta\varphi_1<\pi$. 
The absence of a regular solution in those examples is an indication of a 
divergence, at a finite separation there would be a solution but it 
becomes singular, and the action diverges, as the lines are brought 
together.

The same should generically be true for coincident circles. 
This explains why for most of the range of parameters we will not find 
solutions with finite action. For certain values there are solutions with 
finite action for this case, but one has to take them with a grain of salt, 
we do not know if there is another singular contribution. That will 
require looking at the limit of nearly coincident circles mentioned above.

When the circles are oriented in the same direction there is no such danger, 
the minimal surface has to cover twice the $AdS_2$ subspace (for 
oppositely oriented circles it will cover twice a ring near the boundary of 
$AdS_2$). Still we will find nontrivial solutions only for a small range of 
parameters. The reason is not fully clear to us, but a possible explanation is 
the following intuition from flat space.

Consider two circles of the same orientation in flat space. Generally there 
will be only the disconnected solution, two discs. But if the circles are 
in the same plane and overlapping there is also a connected 
solution. It will cover the same area as the disconnected one, 
but the two sheets will connect there to make 
a single surface. To avoid confusion, the surface that looks like an annulus 
and doesn't cover the smaller circle is not allowed because of the 
orientation.

From this example we see that in generic situations there will not be 
connected solutions for loops of the same orientation. In the case of 
two concentric circles considered in section~\ref{AdS3xS1-section}, 
there is no such solution. On the other hand, if the two surfaces of 
the disconnected solution touch, it's possible to change the topology 
of the surface there into a connected solution.

This indeed happens in this example. If we consider the disconnected 
solution and arrange for both parts to wrap the same hemisphere in 
$S^2$, at the north pole the surface will be at the same position in 
$AdS_5$ (at $r_1=0$). So such a connected solution will exist, and 
will have the same action as the disconnected solution, and 
if the two circles are on the same hemisphere this is the dominant 
solution. This solution has $a^2=0$, exactly like the disconnected one, 
but we may try to look for other connected solutions. We did indeed 
find some such solutions for a certain range of parameters.

Since the coordinate $z$ will not be bound we have to look at the 
$AdS_2$ solution (\ref{coincident-circles-imaginary-a}) with 
$a^2<0$, and the $S^2$ solution (\ref{S2-solution-imaginary-a}). 
We looked in detail at the case when $\theta_i=\theta_f$, where 
the equation for the range of $\sigma$ is
\begin{equation}
\delta\sigma
=\frac{2}{b}\,K\left(\sqrt{1-\frac{k^2}{b^2}}\right)
=\frac{2}{b}\,F\left(\arccos\frac{\cos\theta_i}{\cos\theta_m}
\bigg|\,i\cot\theta_m\right)\,,
\end{equation}
with $b^2=-a^2$ and $\sin\theta_m=b/m$.

For $m>k$ there are no solutions to this equation. To see that, note 
that the right-hand side is not greater than the complete elliptic integral 
$2K(\sqrt{1-m^2/b^2})/b$, which for $m>k$ is smaller than the left-hand 
side. But for all $m<k$ we did find solutions, and the connected solution 
always has smaller action than the disconnected one.

Let us turn now to the case of circles with opposite orientation, 
where the minimal surface will 
not cover the full $AdS_2$ subspace, but stay near the boundary. That 
means the coordinate $z$ will have a maximum, which is the case of 
positive $a^2$. The solution to the $AdS_5$ part of the ansatz is 
given by (\ref{coincident-circles-solution}) and to the $S^5$ ansatz is 
given by (\ref{S2-solution-real-a}).

The range of the world-sheet coordinate $\sigma$ will be given by
(\ref{coincident-circles-delta-sigma}). This has to equal the range calculated 
on the $S^2$ side (\ref{S2-delta-sigma-real-a}), giving the relation
\begin{equation}
\delta\sigma=\frac{2}{a}\,K\left(i\frac{k}{a}\right)
=\frac{1}{a}\left|F\left(\theta_f\bigg|\,i\frac{m}{a}\right)
-F\left(\theta_i\bigg|\,i\frac{m}{a}\right)\right|\,.
\label{coincident-circles-equation}
\end{equation}

Let us assume without loss of generality that $\theta_i<\theta_f$ 
and also that $\theta_i<\pi/2$. 
In the special case that $m=0$ there is no dependence on $\varphi_1$, 
so the $S^5$ ansatz reduces to an $S^1$ and the right hand side becomes 
$(\theta_f-\theta_i)/a$. This is a limiting case of the system 
studied in section~\ref{AdS3xS1-section} when the two circles are 
coincident, i.e. $\delta v=0$. This limit is singular and the action 
diverges.

In general, since $\theta_f\leq\pi$ and $\theta_i\geq0$, the 
right hand side of equation (\ref{coincident-circles-equation}) 
will be smaller than twice the complete elliptic integral 
$2/a\,K(im/a)$.
The complete elliptic integral with imaginary modulus is a 
a monotonously decreasing function from $K(0)=\pi/2$, approaching 
zero at infinite imaginary modulus. Therefore if $m>k$ the right hand 
side is smaller than the left hand side for all values of $a$. Thus there 
will be no regular connected classical solution.

Hence we should study the existence of solutions for $m<k$. 
Let us focus on the case where $\theta_f=\pi-\theta_i$, 
then we use the fact that the elliptic integral with argument 
$\theta_i$ and $\pi-\theta_i$ add up to twice the complete elliptic 
integral to rewrite (\ref{coincident-circles-equation}) as
\begin{equation}
F\left(\theta_i\bigg|\,i\frac{m}{a}\right)
=K\left(i\frac{m}{a}\right)
-K\left(i\frac{k}{a}\right)\,.
\end{equation}
For certain values of $\theta_i$ there will be solutions 
to the equation, and for others not. The solutions will exist for $\theta_i$ 
smaller than a critical value, and will not exist if $\theta_i$ exceeds it. 
We interpret that to mean that when $\theta_i$ and $\theta_f$ are 
too close to each other the phenomenon described above happens---the 
potential between the two circles will diverge.

In all those case there is also a disconnected solution, just two of the 
surfaces described in the preceding subsection. It turns out that in most 
of the range where the connected solution exists, the disconnected one 
has lower action and will dominate. The results of our numerical studies 
are shown in figure~\ref{AdS2xS2-figure}. Connected solutions 
exist to the left of the dashed curve, and they dominate the action only 
in the small region bound by that curve and the solid line.
\begin{figure}[ht]
\centerline{\hbox{\epsffile{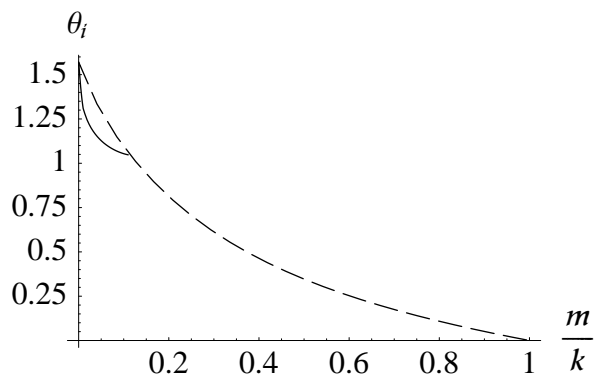}}}
\centerline{\parbox{12cm}{\caption{
\label{AdS2xS2-figure}
Phase diagram for two coincident circles with rotation on $S^2$ at 
angles $\theta_i+\theta_f=\pi$. Regular connected solutions exist only 
to the left of the dashed line, when $\theta_f-\theta_i$ is not too 
small, and $m/k$ not too large. The disconnected solution has smaller 
action in the entire range except for the small region above the solid 
line.
}}}
\end{figure}

\subsection{$AdS_3\times S^3$ subspace}
\label{AdS3xS3-section}

\subsubsection{Anti-parallel lines with rotation}

We wish to consider here two anti-parallel Wilson loops extended in 
the $x_1$ direction, separated by a distance $R$ in the $x_2$ direction 
(\ref{line-loop}). 
Along each of the lines we include in the Wilson loops a coupling to 
the scalars which is periodic around a circle, that is (\ref{S5-loop}) 
with $\psi=\pi/2$ and only $m=m_1\neq0$. So the first line will set the 
initial conditions
\begin{equation}
x_1=\tau\,,\qquad
x_2=-\frac{R}{2}\,,\qquad
\theta=\theta_i\,,\qquad
\varphi_1=m_i\tau\,,\qquad
\varphi_2=\varphi_{2i}\,.
\end{equation}
Along the second line will set the final values
\begin{equation}
x_1=\tau\,,\qquad
x_2=\frac{R}{2}\,,\qquad
\theta=\theta_f\,,\qquad
\varphi_1=m_f\tau\,,\qquad
\varphi_2=\varphi_{2f}\,.
\end{equation}
The $S^3$ ansatz in section~\ref{S3-section} allows the two lines to be 
at different values of $\varphi_2$, as indicated above. In the examples 
we consider below, though, we will take this angle to be a constant, to 
simplify the expressions, which will then fit the $S^2$ ansatz in section 
\ref{S2-section}.

In the above ansatz we allowed the two lines to rotate with different 
parameters $m$. If $m_f=m_i$ the two lines rotate together, while for 
$m_f=-m_i$, they are rotating in opposite directions. In the former case 
the angle $\theta$ will not have to go through zero. In the latter, since the 
two rotations have opposite orientation, $\theta$ will have to go through 
zero. In fact, we can consider this case on the same footing as the 
other by the replacement $m_f\to -m_f$ and $\theta_f\to-\theta_f$. 
So negative $\theta$ correspond to rotation with the opposite 
orientation.

Another case we may consider is for $m_f=0$, which is realized by 
$\theta_f=0$.

Some of those loops were already studied by Tseytlin and Zarembo 
\cite{Tseytlin:2002tr}. We generalize their solutions and study them 
further.

As usual, we will have to equate the range of the world-sheet coordinate 
$\sigma$, which for the $S^3$ ansatz can be found from the expressions 
in section~\ref{S3-section}. Since we restrict ourselves to the $S^2$ case 
we have (\ref{S2-delta-sigma-real-a}) for $a^2>0$ and 
(\ref{S2-delta-sigma-imaginary-a}) for $a^2<0$. From the $AdS_5$ part 
of the ansatz $\delta\sigma$ is given by (\ref{lines-delta-sigma}).

Let us first study the case where $m_f=m_i$ (and label it $m$). The 
initial and final values $\theta_i$ and $\theta_f$ are both positive. We 
found four different types of solutions for these boundary conditions 
that will be realized for different values of the parameters. We 
illustrate some of them in figure~\ref{AdS3xS3-3-figure} for 
$\theta_i=5\pi/12$ and $\theta_f=2\pi/3$.
\begin{figure}[ht]
\centerline{\hbox{\epsffile{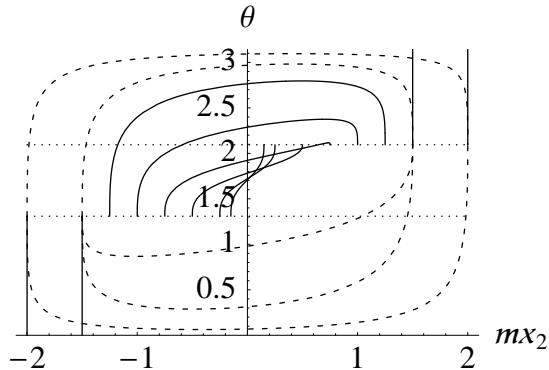}}}
\centerline{\parbox{12cm}{\caption{
\label{AdS3xS3-3-figure}
The values of $\theta$ as a function of the spatial coordinate $x_2$ 
for boundary values $\theta_i=5\pi/12$ and $\theta_f=2\pi/3$ 
 (the dotted horizontal lines) and different separation. At very large 
separation ($mR=3,\,4$) there are connected solutions (dashed) but 
the disconnected solution (solid) dominates. As shorter separation 
($mR=1.5,\,2,\,2.5$) the connected solutions dominate and 
have a turning point $\theta_m>\theta_f$. At shorter distances 
the solution does not have a turning point.
}}}
\end{figure}

One solution, which exists for all values of the parameters is the 
disconnected one. It is simply two of the lines considered in 
section~\ref{AdS2xS2-line}, with action
\begin{equation}
\cS=\frac{T\sqrt\lambda}{2\pi}\,m
(2-|\cos\theta_i|-|\cos\theta_f|)\,.
\end{equation}

There are also connected solutions for all values of the parameters, 
but their nature depends on the separation between the lines. When the 
two lines are very far apart the connected solution will be described by 
the solution with negative $a^2$ (\ref{S2-solution-imaginary-a}) with 
a turning point $\theta_m$. It will start at $\theta_i$, go to 
$\theta_m$, near $0$ or $\pi$ and then back 
to $\theta_f$. For some values there would be only one such solution, 
and in other cases two or even more (see figure~\ref{AdS3xS3-3-figure}).

If the distance between the lines is large enough it's also possible to 
construct classical solutions that oscillate a few times. They will 
go to some $\theta_m$, then to $\pi-\theta_m$, and back. Those 
solutions will never dominate the action.

As the lines get closer, $\theta_m$ will approach $\theta_i$ or 
$\theta_f$, and beyond that this branch of the solution will cease to 
exist. Instead there will be a new solution which still has negative 
$a^2$, but will not have a turning point. Those solutions can still be 
described by some value of $\theta_m$, but it will not be along the 
world-sheet. As the lines get closer this $\theta_m$ will get again 
closer to $0$ or $\pi$.

Finally, when $\theta_m$ reaches the north or south pole we have to 
look at solutions with $a=0$, which are quite easy to study. 
In this special case the solution to the $AdS_5$ part is 
identical to that of the parallel lines with no motion on $S^5$, 
studied in \cite{rey-wl,maldacena-wl} and reviewed in 
section~\ref{AdS3xS1-section}. The separation between the lines 
is given by (\ref{AdS3xS1-R}) as a function of $p$. The value of 
$p$ can be found by solving the equation for the range of $\sigma$, 
which using (\ref{S2-zero-a}) is
\begin{equation}
\delta\sigma=
\frac{2}{\sqrt{p}}\,K(i)
=\frac{\Gamma(1/4)^2}{2\sqrt{2\pi}}\frac{1}{\sqrt{p}}
=\frac{1}{m}\left[\hbox{arccosh}\frac{1}{\sin\theta_i}
+\hbox{arccosh}\frac{1}{\sin\theta_f}\right]\,.
\label{AdS3xS3-zero-a}
\end{equation}

At even closer separation $a^2$ will be positive and the solution will 
be described by (\ref{S2-solution-real-a}).

This suggests a phase diagram with four phases, one where the 
disconnected solution dominates, and three with connected solutions: 
Negative $a^2$ and a turning point, negative $a^2$ without a turning 
point and positive $a^2$. We now turn to study those phases in some 
specific examples. 

Let's start considering the particular case $\theta_i=\theta_f$. 
In this simple case we find that almost always $a^2<0$, and there 
is a turning point at
$\sin\theta_m=b/m$ with $b^2=-a^2$. In that case we find the 
relation (\ref{S2-delta-sigma-imaginary-a}), (\ref{lines-delta-sigma})
\begin{equation}
\delta\sigma=2y_+
K\left(ipy_+^2\right)
=\frac{2}{b}F\left(\arccos\frac{\cos\theta_i}{\cos\theta_m}
\bigg|\,i\cot\theta_m\right)\,.
\label{AdS3xS3-delta-sigma-1}
\end{equation}
with (\ref{lines-y+})
\begin{equation}
y_+^2=\frac{b^2+\sqrt{b^4+4p^2}}{2p^2}\,.
\end{equation}

For $\theta_i<\pi/2$ we can always find a solution to these equations. 
The further the two lines are from each other the smaller $\sin\theta_m$ 
would be, and as they get closer $\theta_m$ approaches the initial 
value $\theta_i$. In the special case of $\theta_i=\pi/2$, were the 
initial and final positions are along the equator of $S^2$ the solution 
ceases to exist for $mR\lesssim2.312$. At shorter distances 
we have to use the other solution discussed at the end of 
section~\ref{S2-section}, where $\theta_m=\pi/2$. Without the 
$AdS_5$ contribution this solution would be unstable, but here it is 
in fact realized. The transition between those two solutions was 
explained in \cite{Tseytlin:2002tr}, which studied exactly this system 
with $\theta_i=\theta_f=\pi/2$.

We illustrate this behavior in figure~\ref{AdS3xS3-1-figure}, where 
we plot the value of the turning point $\theta_m$ as a function of the 
distance between the lines (multiplied by the rotation parameter $m$, to 
make it scale invariant) for different values of $\theta_i$. 
As $mR\to0$ the turning point approaches $\theta_i$, and it gets 
closer to the pole as $mR$ is increased. In the special case of 
$\theta_i=\pi/2$ the curve is non-differentiable, due to the phase 
transition between the two solutions. For all other values of $\theta_i$ 
the curve is smooth, but the bend gets sharper as we approach 
$\theta_i=\pi/2$.
\begin{figure}[ht]
\centerline{\hbox{\epsffile{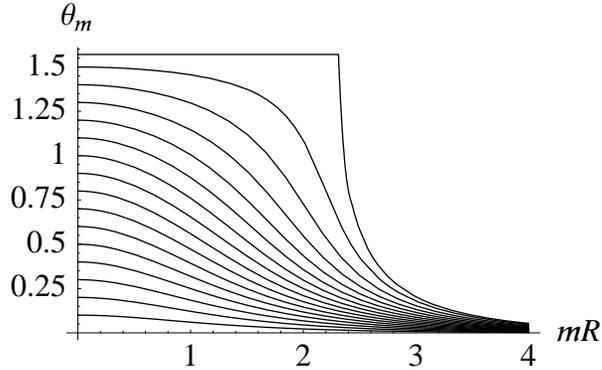}}}
\centerline{\parbox{12cm}{\caption{
\label{AdS3xS3-1-figure}
$\theta_m$, the minimal value of $\theta$ for two lines 
separated a distance $R$ with wrapping $m$ times around the sphere. 
The initial and final values of $\theta$ 
(which are equal in this case) can be read from the intersect of the 
lines with the axis at $R=0$, since then $\theta_m=\theta_i$. 
The bend in the curve gets sharper the larger $\sin\theta_i$ is, 
and for $\theta_i=\pi/2$ it is not differentiable.
}}}
\end{figure}

The action for this solution is given by the sum of the $S^2$ contribution 
(\ref{S2-action-imaginary-a}) and that of $AdS_5$ 
(\ref{lines-action}). It is always smaller than that of the disconnected 
solution.

Another example where the expressions are simple, but the phase structure 
much richer is for $\theta_f=\pi-\theta_i$. At very large separation 
the disconnected solution will always dominate, the two pieces will 
cover parts of the two hemispheres, and will not cross $\theta=\pi/2$ 
as the connected solution has to. The action is (\ref{S2-action-zero-a})
\begin{equation}
\cS=\frac{T\sqrt\lambda}{2\pi}\,
2m(1-|\cos\theta_i|)\,.
\label{AdS3xS3-action-disconnected}
\end{equation}

This solution will coexist with the connected solutions and dominate 
at large distances. There will be a first order phase transition 
when the connected solutions start dominating. 
Figure~\ref{AdS3xS3-2-figure} shows the different phases for those 
boundary values, the disconnected solutions dominate everywhere to the 
right of the solid line.

\begin{figure}[ht]
\centerline{\hbox{\epsffile{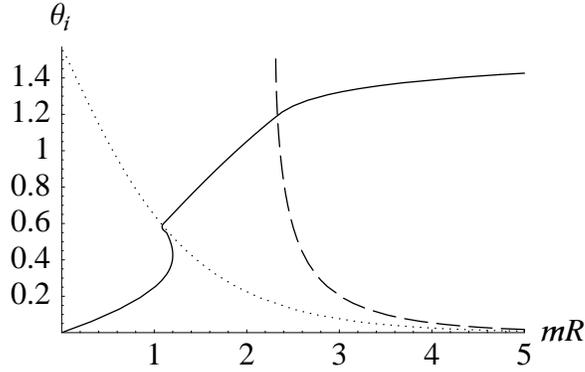}}}
\centerline{\parbox{12cm}{\caption{
\label{AdS3xS3-2-figure}
The phase diagram for two lines wrapping the sphere at angles 
$\theta_i+\theta_f=\pi$. 
For all values of $\theta_i$ and $R$ there are at least two solutions, 
including one disconnected and one connected. The disconnected 
solution has lower action to the right of the solid curve, and the 
connected ones to the left. The connected solution has positive 
$a^2$ to the left of the dotted curve (given by (\ref{AdS3xS3-zero-a-2})) 
Between the dotted and dashed curve the solution has negative 
$a^2$, but no turning point. And to the right of the dashed line the 
solution has a turning point. The solid line and dotted one seem to 
represent first order phase transitions, and the dashed one a second 
order one.
}}}
\end{figure}

For large separation $R$ the connected solution that is realized is the 
one with a turning point at $\sin\theta_m=b/m$. Since $\theta_i$ 
and $\theta_f$ are complementary the sum of the two elliptic integrals 
in (\ref{S2-delta-sigma-imaginary-a}) is the complete integral. 
Combining with (\ref{lines-delta-sigma}) we get the relation
\begin{equation}
\delta\sigma
=2y_+\,K\left(ipy_+^2\right)
=\frac{2}{b}\,K\left(i\cot\theta_m\right)\,,
\end{equation}
where $y_+$ is like in the previous example. Note that this equation 
is independent of the value of $\theta_i$, and is identical to the 
equation in the last example (\ref{AdS3xS3-delta-sigma-1}) for the 
case of $\theta_i=\theta_f=\pi/2$.

This equation can always be solved, but the solution may have 
$\sin\theta_m=b/m>\sin\theta_i$, which is unphysical. The 
value of $\theta_m$ as a function of $mR$ is the uppermost curve 
in figure~\ref{AdS3xS3-1-figure}. In the present case this phase 
ceases to exist for $\theta_i=\theta_m$. Therefore this same curve 
serves now as the phase boundary and is illustrated by the dashed line 
in figure~\ref{AdS3xS3-2-figure}.

Beyond that point the connected solution will still have negative 
$a^2$, but there will not be a turning point.  It will be described 
by the (\ref{S2-solution-imaginary-a}) with only the positive 
branch and the negative signs in (\ref{S2-delta-sigma-imaginary-a})
and (\ref{S2-action-imaginary-a}). The range of $\sigma$ is
\begin{equation}
\delta\sigma
=2y_+\,K\left(ipy_+^2\right)
=\frac{2}{b}\left[
K\left(i\cot\theta_m\right)
-F\left(\arccos\frac{\cos\theta_i}{\cos\theta_m}\bigg|\,
i\cot\theta_m\right)\right]\,.
\end{equation}
The action for this solution is given by the usual expressions. It 
is interesting to look at it close to the transition to the phase discussed 
before. From the numerical data it seems like the phase transition is 
of second order.

As the separation $R$ is decreased we find that $b\to0$, which happens 
at a value of $p$ given by (\ref{AdS3xS3-zero-a}). Combining that 
with (\ref{AdS3xS1-R}) we find the relation between $R$ and 
$\theta_i$ to be
\begin{equation}
\cosh\left(\frac{\Gamma(1/4)^4}{16\pi^2}mR\right)
=\frac{1}{\sin \theta_i}\,.
\label{AdS3xS3-zero-a-2}
\end{equation}
This relation is shown by the dotted line in figure~\ref{AdS3xS3-2-figure}.

At shorter separations the connected solution will have positive $a^2$. 
Numerical analysis suggests the phase transition between negative and 
positive $a^2$ is of first order.

Note that for the special case of $\theta_i=0$ and $\theta_f=\pi$ 
the string starts at the north pole and ends at the south pole. In this case 
$m$ plays no role at the boundary. In fact this configuration has the 
same boundary conditions as for two lines in the $AdS_3\times S^1$ 
ansatz separated by an angle $\delta\varphi=\pi$. In 
section~\ref{AdS3xS1-section} we saw that in this case there was no 
connected solution. Allowing for the $S^2$ ansatz here, we do find 
a connected solution, but its action is greater than the disconnected one 
for all separations (the bottom of figure~\ref{AdS3xS3-2-figure}).

Let us now leave this case and look briefly at some examples with 
$m_f=-m_i$. It is simpler, as explained above, to describe them by 
allowing negative $\theta$, so we take positive $\theta_i$ and negative 
$\theta_f$ and replace $m_f=m_i=m$.

In this case there seem to be only two phases, the connected solution 
with $a^2>0$ and a disconnected one. The disconnected one will 
not be dominant unless $\theta_i$ and $\theta_f$ are on different 
hemispheres.

Thus an interesting case is the one with $\theta_f=\theta_i-\pi$ 
(or alternatively $\theta_f=\pi-\theta_i$ and $m_f=-m_i$). The 
connected solution has $a^2>0$, and again the sum of the two elliptic 
integrals is the complete integral
\begin{equation}
\delta\sigma=2y_+
K\left(ipy_+^2\right)
=\frac{2}{a}\,K\left(i\frac{m}{a}\right)\,.
\end{equation}

At short distances this connected solution will dominate, 
but at large distances the action of the disconnected solution, given in 
(\ref{AdS3xS3-action-disconnected}) will be lower. The line separating 
those two phases is shown in figure~\ref{AdS3xS3-4-figure}.
\begin{figure}[ht]
\centerline{\hbox{\epsffile{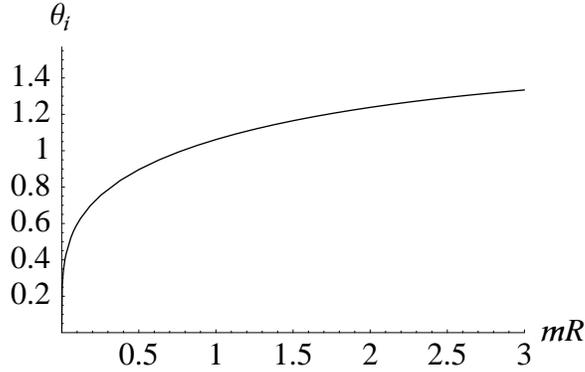}}}
\centerline{\parbox{12cm}{\caption{
\label{AdS3xS3-4-figure}
The phase diagram for two lines with rotations in opposite directions 
at angles $\theta_f=\theta_i-\pi$. 
For all values of the separation $R$ there are connected solutions 
with $a^2>0$, but to the right of the curve the disconnected solution 
dominates.
}}}
\end{figure}

Finally let us comment on the case where $m_f=0$. That is easy to 
incorporate into this ansatz by taking $\theta_f=0$ (or $\theta_f=\pi$). 
In this case we find a classical solution with $a^2>0$ for all values of 
$m$ and $\theta_i$, and it always has smaller action than the disconnected 
solution.

\subsubsection{Concentric circles with rotation}

Another periodic ansatz that fits inside the $AdS_3$ subspace of 
$AdS_5$ is the case of two concentric circles. We now consider this 
example where the Wilson loop also couples to the scalars in a periodic 
fashion. The boundary conditions on the first circle are
\begin{equation}
v=v_i=\log R_i\,,\qquad
\phi_1=k\tau\,,\qquad
\theta=\theta_i\,,\qquad
\varphi_1=m_i\tau\,,\qquad
\varphi_2=\varphi_{2i}\,.
\end{equation}
For the second circle we have
\begin{equation}
v=v_i=\log R_i\,,\qquad
\phi_1=k\tau\,,\qquad
\theta=\theta_f\,,\qquad
\varphi_1=m_f\tau\,,\qquad
\varphi_2=\varphi_{2f}\,.
\end{equation}
We will not study this general ansatz in detail, there are many parameters to 
vary and presumably many different phases. We will look only at the case 
with $\theta_i=\theta_f=\pi/2$, $m_f=m_i=m$ and 
$\varphi_{2f}=\varphi_{2i}$.

For these values of $\theta_i$ and $\theta_f$ there are two possible 
solutions to the $S^2$ ansatz, that with $-m^2<a^2<0$ 
(\ref{S2-solution-imaginary-a}) and the one with $a^2=-m^2$ mentioned 
at the end of section~\ref{S2-section} (and was realized also in the 
last subsection).

For the case when $a^2<0$, the equations for the range of $\sigma$ 
are (\ref{S2-delta-sigma-imaginary-a}), 
(\ref{two-circles-delta-sigma}) 
\begin{equation}
\delta\sigma=\frac{2z_+}{k}
K\left(\frac{z_+}{z_-}\right)
=\frac{2}{b}\,K\left(i\cot\theta_m\right)\,.
\end{equation}
with $b^2=-a^2$ and (\ref{circles-nu})
\begin{equation}
z_\pm^2=\frac{b^2+k^2\pm\sqrt{(b^2-k^2)^2+4k^2p^2}}{2(p^2-b^2)}\,,
\end{equation}

\begin{figure}[ht]
\centerline{\hbox{\epsffile{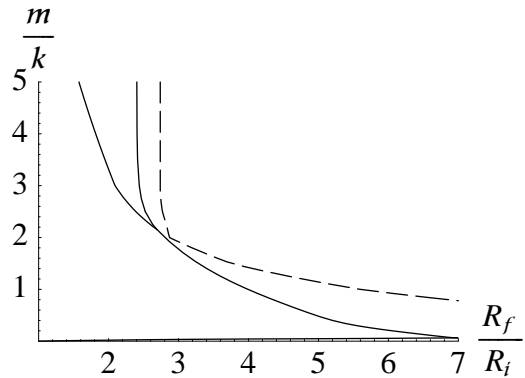}}}
\centerline{\parbox{12cm}{\caption{
\label{AdS3xS3-c1-figure}
Phase diagram for concentric circles with radii $R_f>R_i$ which rotate 
$k$ times in $\bR^4$ and $m$ times on the sphere with 
$\theta_i=\theta_f=\pi/2$. The disconnected solution as well as the 
solution with constant $\theta$ exist for all values of the parameters, 
and in addition there are solutions with a turning point to the left of the 
dashed line. The solid curve is the phase boundary between that 
dominated by the disconnected solution (to the right), the connected 
solution with constant $\theta$ (to the left) and the solution with a 
turning point (above the triple point).
}}}
\end{figure}

The results of the numerical studies are shown in 
figure~\ref{AdS3xS3-c1-figure}. This figure is for $R_f>R_i$, and 
can be extended by the replacement $R_f\leftrightarrow R_i$. 
For all values of $m$, $k$, $R_f$ and $R_i$ both the disconnected solution 
and the connected one with constant $\theta=\pi/2$ exist. The disconnected 
one dominates for large values of the ratio $R_f/R_i$, to the right of the 
solid curve. What is not obvious from the picture is that $R_f/R_i$ has 
a maximum of roughly $6.9914$ at small $m/k$ and beyond that the 
curve goes to $R_f/R_i=1$ at $m/k=0$.

The solution with negative $a^2$ exists to the left of the dashed line 
and dominates the action above the triple point (note that while the dashed 
line gets very close to the solid one, they do not touch). For very large 
$m/k$ the leftmost solid line approaches 1, so the unstable 
solution does not dominate there. The other two lines approach the values 
$2.4034$ and $2.7245$, which are the same as the intercept of the lines with 
the horizontal axis in fig.~\ref{AdS3xS1-figure}. That was the case of 
two concentric circles with no motion on the $S^5$ (also studied by 
\cite{Zarembo:1999bu,Olesen:2000ji}).

\section{Discussion}

The purpose of this article is twofold; to point out that the integrability 
of the $\sigma$-model on $AdS_5\times S^5$ can be used to calculate 
Wilson loops within the $AdS$/CFT correspondence, and to demonstrate 
this in some specific examples.

It is very easy to integrate the equations of motion, as we did in 
sections~\ref{S5-section} and~\ref{AdS5-section}. The results are 
given by trigonometric and hyperbolic functions as well as elliptic 
integrals, and more complicated cases would be 
expressed in terms of hyperelliptic integrals. On the other hand, 
combining the solutions into full-fledged examples leads to very complex 
phenomena. There are many solutions to the equations of motion for 
the same boundary conditions and it requires careful study to find which 
will dominate.

Some of the solutions exist only for a certain range of parameters, like 
the catenoid discussed in 
\cite{Gross:1998gk,Zarembo:1999bu,Olesen:2000ji}. Other solutions 
may coexist for the entire range of parameters (like the connected and 
disconnected solutions for anti-parallel lines with rotation), with one 
dominating in a certain regime, and the other in another.

Thus the equations of motion are easy to solve, but fixing the boundary 
conditions leads to the complicated phase structure. This, in fact, is not 
unlike the classical problem of soap films in flat space (``Plateau's 
problem''). The difficulty is not with solving the equations of motion 
locally, but with satisfying the boundary conditions. These complications 
raise some questions we would like to discuss now.

Some recent studied of local operators in $\cN=4$ supersymmetric 
Yang-Mills seem to indicate that in the planar limit the theory is 
integrable. The integrable structure was found both on the field 
theory side at weak coupling and in $AdS_5\times S^5$, corresponding 
to large 't Hooft coupling. In fact, integrability in this limit is 
based exactly on the same integrals of motion considered here for 
the Wilson loop operators.

If the full planar theory is to be integrable, it should include also 
the Wilson loop operators. As stated, the equations of motion are 
integrable, which agrees with those expectations, but the existence of 
multiple solutions complicates matters and is a question that will have 
to be addressed in making the claim that the planar theory is integrable.

Many confusing issues arise when trying to compare what is known 
about the integrability of local operators to Wilson loops. Are the 
basic objects the one-point functions of those operators, or the 
correlators of two? What is the meaning of the different 
phase transitions, where the correlator of two loops changes 
markedly as the boundary conditions are modified.

To gain further intuition into those problems it is useful to find 
some simple operators that can be used as starting points in this 
investigation. Here we tried to classify the loops according to 
subspaces they reside within. Clearly the simplest is the circle or 
line, which live inside an $AdS_2$ subspace of $AdS_5\times S^5$. 
The next level of complication comes from extending it either 
to $AdS_3\times S^1$ or to $AdS_2\times S^2$. Beyond that 
we considered the case of $AdS_3\times S^3$.

This classification has some resemblance to the subsectors of local 
operators related to different spin-chains. The Wilson loops whose 
minimal surface fit within the $AdS_3\times S^1$ include two 
components of the gauge field and two real scalars. That is 
reminiscent of the spin-chain with one complex scalar and a chiral 
combination of covariant derivatives. The operators described within 
$AdS_2\times S^2$ have a single gauge field and three scalars, 
which can also be combined into a complex scalar and a chiral 
operator made of a covariant derivative and scalar. Finally 
the $AdS_3\times S^3$ subspace is related to operators with 
two covariant derivatives and four real scalars, which again 
has some similarity to the sector of local operators made of 
two complex scalars and a chiral covariant derivative.

In this paper we studied Wilson loops only in $AdS_5\times S^5$, 
and did not touch at all on the gauge theory side of the duality. While 
the periodic ansatze led to simple equations for the string surface which 
we solved, the perturbative calculation seems quite complicated. We 
did not find analogous signs of integrable structure in the weakly 
coupled field theory.

There are also some conceptual problems that have to be addressed in 
this comparison. In $AdS_5\times S^5$ one can consider operators 
that are localized at the same point on the boundary of $AdS_5$ but 
separated only in the $S^5$ direction. For example, in the example of 
the single circle in section~\ref{AdS2xS2-circle-section} we considered 
the circle wrapping the curve in space $k$ times while wrapping a 
circle in $S^5$ $m$ times. There is a nice minimal surfaces with 
finite action for all integers values of $k$ and $m$. If one looks at 
the same Wilson loop operator in the gauge theory, there will be 
singular graphs that connect different points along the loop 
that are coincident in space. Those graphs will lead to divergences unless 
$m$ is divisible by $k$. 

A more extreme example of this is for coincident circles (see 
section~\ref{coincident-circles-section}), where for certain values 
of the parameters we found finite action solutions. Calculating 
those Wilson loops in perturbation theory involves extremely 
divergent graphs.

This issue has some similarities to the question of zig-zag symmetry. 
We know that a Wilson loop does not change under arbitrary 
reparametrizations, including backtracking. But looking at the part 
of the loop that backtracks naively in perturbation theory gives a 
very bad divergence. The coincident circles resemble a backtracking 
curve, only with different couplings to the scalars, and to capture them 
requires some regularization that satisfies this general form of 
zig-zag symmetry allowing for extra motion on $S^5$.

Another difference between the string theory calculation and that on 
the gauge theory side is that in the gauge theory at weak coupling there 
is no sign of those complicated phases we find for the minimal 
surfaces. This is a phenomenon that must show up only at the level 
of classical string theory, but the phase transitions should be smoothed 
out by quantum corrections to the $\sigma$-model, and are hard to 
see from the gauge theory side. There actually is an example of a 
class of operators where by summing the perturbative result and 
expressing the results in a $1/\sqrt\lambda$ expansion, one finds 
two saddle points that agree with two solutions found in string theory
\cite{we}.

\section*{Acknowledgments}
We would like to thank Gordon Semenoff and Matthias Staudacher 
for useful discussions.
N.D. would like to thank the Weizmann institute, where this project was 
initiated, for its hospitality.
The authors would like to thank the Aspen Center for physics for 
providing the opportunity to 
continue this collaboration in an inspiring atmosphere.


\begin{thebibliography}{20}
%%%%%%%%%%%%%%%%%%%%%%%%%%%
\addtolength{\parskip}{-1ex}

\bibitem{ads/cft}
J.~M.~Maldacena,
``The large~$N$ limit of superconformal field theories and supergravity,''
Adv.\ Theor.\ Math.\ Phys.\  {\bf 2}, 231 (1998)
[Int.\ J.\ Theor.\ Phys.\  {\bf 38}, 1113 (1999)]
[arXiv:hep-th/9711200].
%%CITATION = HEP-TH 9711200;%%

\bibitem{Beisert:2004ry}
N.~Beisert,
``The dilatation operator of $\cN = 4$ super Yang-Mills theory 
and integrability,''
Phys.\ Rept.\  {\bf 405}, 1 (2005)
[arXiv:hep-th/0407277].
%%CITATION = HEP-TH 0407277;%%

\bibitem{Pohlmeyer:1975nb}
K.~Pohlmeyer,
``Integrable Hamiltonian Systems And Interactions Through Quadratic
Constraints,'' Commun.\ Math.\ Phys.\  {\bf 46} (1976) 207.
%%CITATION = CMPHA,46,207;%%
M.~Luscher and K.~Pohlmeyer,
``Scattering Of Massless Lumps And Nonlocal Charges In The Two-Dimensional
Classical Nonlinear Sigma Model,''
Nucl.\ Phys.\ B {\bf 137}, 46 (1978).
%%CITATION = NUPHA,B137,46;%%

\bibitem{Metsaev:1998it}
R.~R.~Metsaev and A.~A.~Tseytlin,
``Type IIB superstring action in $AdS_5\times S^5$ background,''
Nucl.\ Phys.\ B {\bf 533}, 109 (1998)
[arXiv:hep-th/9805028].
%%CITATION = HEP-TH 9805028;%%

\bibitem{Mandal:2002fs}
G.~Mandal, N.~V.~Suryanarayana and S.~R.~Wadia,
``Aspects of semiclassical strings in $AdS_5$,''
Phys.\ Lett.\ B {\bf 543}, 81 (2002)
[arXiv:hep-th/0206103].
%%CITATION = HEP-TH 0206103;%%

\bibitem{Bena:2003wd}
I.~Bena, J.~Polchinski and R.~Roiban,
``Hidden symmetries of the $AdS_5\times S^5$ superstring,''
Phys.\ Rev.\ D {\bf 69}, 046002 (2004)
[arXiv:hep-th/0305116].
%%CITATION = HEP-TH 0305116;%%

\bibitem{Gubser:2002tv}
S.~S.~Gubser, I.~R.~Klebanov and A.~M.~Polyakov,
``A semi-classical limit of the gauge/string correspondence,''
Nucl.\ Phys.\ B {\bf 636}, 99 (2002)
[arXiv:hep-th/0204051].
%%CITATION = HEP-TH 0204051;%%

\bibitem{Frolov:2002av}
S.~Frolov and A.~A.~Tseytlin,
``Semiclassical quantization of rotating superstring in $AdS_5\times S^5$,''
JHEP {\bf 0206}, 007 (2002)
[arXiv:hep-th/0204226].
%%CITATION = HEP-TH 0204226;%%

\bibitem{Arutyunov:2003uj}
G.~Arutyunov, S.~Frolov, J.~Russo and A.~A.~Tseytlin,
``Spinning strings in $AdS_5\times S^5$ and integrable systems,''
Nucl.\ Phys.\ B {\bf 671}, 3 (2003)
[arXiv:hep-th/0307191].
%%CITATION = HEP-TH 0307191;%%

\bibitem{Arutyunov:2003za}
G.~Arutyunov, J.~Russo and A.~A.~Tseytlin,
``Spinning strings in $AdS_5\times S^5$: New integrable system relations,''
Phys.\ Rev.\ D {\bf 69}, 086009 (2004)
[arXiv:hep-th/0311004].
%%CITATION = HEP-TH 0311004;%%

\bibitem{Minahan:2002ve}
J.~A.~Minahan and K.~Zarembo,
``The Bethe-ansatz for $\cN = 4$ super Yang-Mills,''
JHEP {\bf 0303}, 013 (2003)
[arXiv:hep-th/0212208].
%%CITATION = HEP-TH 0212208;%%

\bibitem{Berenstein:1999ij}
D.~Berenstein, R.~Corrado, W.~Fischler and J.~Maldacena,
``The operator product expansion for Wilson loops and surfaces in the
large~$N$ limit,''
Phys.\ Rev.\  {\bf D59}, 105023 (1999)
[hep-th/9809188].
%%CITATION = HEP-TH 9809188;%%

\bibitem{Drukker:1999zq}
N.~Drukker, D.~J.~Gross and H.~Ooguri,
``Wilson Loops and Minimal Surfaces,''
Phys.\ Rev.\  {\bf D60}, 125006 (1999)
[hep-th/9904191].
%%CITATION = HEP-TH 9904191;%%

\bibitem{erickson}
J.~K.~Erickson, G.~W.~Semenoff and K.~Zarembo,
``Wilson loops in $\cN = 4$ supersymmetric Yang-Mills theory,''
Nucl.\ Phys.\ B {\bf 582}, 155 (2000)
[arXiv:hep-th/0003055].
%%CITATION = HEP-TH 0003055;%%

\bibitem{Drukker:2000rr}
N.~Drukker and D.~J.~Gross,
``An exact prediction of $\cN = 4$ SUSYM theory for string theory,''
J.\ Math.\ Phys.\  {\bf 42}, 2896 (2001)
[arXiv:hep-th/0010274].
%%CITATION = HEP-TH 0010274;%%

\bibitem{Drukker:2005kx}
N.~Drukker and B.~Fiol,
``All-genus calculation of Wilson loops using D-branes,''
JHEP {\bf 0502}, 010 (2005)
[arXiv:hep-th/0501109].
%%CITATION = HEP-TH 0501109;%%

\bibitem{Semenoff:2001xp}
G.~W.~Semenoff and K.~Zarembo,
``More exact predictions of SUSYM for string theory,''
Nucl.\ Phys.\ B {\bf 616}, 34 (2001)
[arXiv:hep-th/0106015].
%%CITATION = HEP-TH 0106015;%%

\bibitem{Zarembo:2002ph}
K.~Zarembo,
``Open string fluctuations in $AdS_5\times S^5$ and operators with large 
$R$ charge,''
Phys.\ Rev.\ D {\bf 66}, 105021 (2002)
[arXiv:hep-th/0209095].
%%CITATION = HEP-TH 0209095;%%

\bibitem{rey-wl}
S.~J.~Rey and J.~T.~Yee,
``Macroscopic strings as heavy quarks in large~$N$ gauge theory and anti-de
Sitter supergravity,''
Eur.\ Phys.\ J.\ C {\bf 22}, 379 (2001)
[arXiv:hep-th/9803001].
%%CITATION = HEP-TH 9803001;%%

\bibitem{maldacena-wl}
J.~M.~Maldacena,
``Wilson loops in large~$N$ field theories,''
Phys.\ Rev.\ Lett.\  {\bf 80}, 4859 (1998)
[arXiv:hep-th/9803002].
%%CITATION = HEP-TH 9803002;%%

\bibitem{Zarembo:1999bu}
K.~Zarembo,
``Wilson loop correlator in the $AdS$/CFT correspondence,''
Phys.\ Lett.\ B {\bf 459}, 527 (1999)
[arXiv:hep-th/9904149].
%%CITATION = HEP-TH 9904149;%%

\bibitem{Olesen:2000ji}
P.~Olesen and K.~Zarembo,
``Phase transition in Wilson loop correlator from $AdS$/CFT 
correspondence,''
arXiv:hep-th/0009210.
%%CITATION = HEP-TH 0009210;%%

\bibitem{Zarembo:2002an}
K.~Zarembo,
``Supersymmetric Wilson loops,''
Nucl.\ Phys.\ B {\bf 643}, 157 (2002)
[arXiv:hep-th/0205160].
%%CITATION = HEP-TH 0205160;%%

\bibitem{Tseytlin:2002tr}
A.~A.~Tseytlin and K.~Zarembo,
``Wilson loops in $\cN = 4$ SYM theory: Rotation in $S^5$,''
Phys.\ Rev.\ D {\bf 66}, 125010 (2002)
[arXiv:hep-th/0207241].
%%CITATION = HEP-TH 0207241;%%

\bibitem{Tseytlin:2003ii}
A.~A.~Tseytlin,
``Spinning strings and $AdS$/CFT duality,''
arXiv:hep-th/0311139.
%%CITATION = HEP-TH 0311139;%%

\bibitem{Babelon:1992rb}
O.~Babelon and M.~Talon,
``Separation of variables for the classical and quantum Neumann model,''
Nucl.\ Phys.\ B {\bf 379}, 321 (1992)
[arXiv:hep-th/9201035].
%%CITATION = HEP-TH 9201035;%%

\bibitem{Mikhailov:2002ya}
A.~Mikhailov,
``Special contact Wilson loops,''
arXiv:hep-th/0211229.
%%CITATION = HEP-TH 0211229;%%

\bibitem{Bianchi:2002gz}
M.~Bianchi, M.~B.~Green and S.~Kovacs,
``Instanton corrections to circular Wilson loops in $\cN = 4$ supersymmetric
Yang-Mills,''
JHEP {\bf 0204}, 040 (2002)
[arXiv:hep-th/0202003].
%%CITATION = HEP-TH 0202003;%%

\bibitem{Staudacher:1997kn}
M.~Staudacher and W.~Krauth,
``Two-dimensional QCD in the Wu-Mandelstam-Leibbrandt prescription,''
Phys.\ Rev.\ D {\bf 57}, 2456 (1998)
[arXiv:hep-th/9709101].
%%CITATION = HEP-TH 9709101;%%

\bibitem{Gross:1998gk}
D.~J.~Gross and H.~Ooguri,
``Aspects of large~$N$ gauge theory dynamics as seen by string theory,''
Phys.\ Rev.\ D {\bf 58}, 106002 (1998)
[arXiv:hep-th/9805129].
%%CITATION = HEP-TH 9805129;%%

\bibitem{we}
N.~Drukker and B.~Fiol, In preparation.

\bibitem{Guralnik:2003di}
Z.~Guralnik and B.~Kulik,
``Properties of chiral Wilson loops,''
JHEP {\bf 0401}, 065 (2004)
[arXiv:hep-th/0309118].
%%CITATION = HEP-TH 0309118;%%

\bibitem{Guralnik:2004yc}
Z.~Guralnik, S.~Kovacs and B.~Kulik,
``Less is more: Non-renormalization theorems from lower dimensional
superspace,''
arXiv:hep-th/0409091.
%%CITATION = HEP-TH 0409091;%%

\end{thebibliography}
\end{document}